\documentclass[10pt,journal,onecolumn]{IEEEtran}
\usepackage{amssymb}
\usepackage{graphicx}
\usepackage{amsmath}
\usepackage{epsf}
\usepackage{mathrsfs}

\newtheorem{theorem}{Theorem}

\newtheorem{definition}{Definition}
\newtheorem{corollary}{Corollary}

\newtheorem{remark}{Remark}
\newtheorem{lemma}{Lemma}

\begin{document}
%

\title{Capacity Bounds for \\ the Gaussian Interference Channel}
\author{\Large Abolfazl~S.~Motahari,~\IEEEmembership{Student Member,~IEEE,}
        and~Amir~K.~Khandani,~\IEEEmembership{Member,~IEEE}
\\\small Coding \& Signal Transmission Laboratory (www.cst.uwaterloo.ca)
\\ \{abolfazl,khandani\}@cst.uwaterloo.ca
}


\maketitle

\footnotetext[1]{\textbf{An earlier version of this work containing
all the results is reported in {\em Library and Archives Canada
Technical Report \underline{UW-ECE 2007-26}, Aug. 2007} (see
http://www.cst.uwaterloo.ca/pub\_tech\_rep.html for details).}}

\begin{abstract}
The capacity region of the two-user Gaussian Interference Channel
(IC) is studied. Three classes of channels are considered: weak,
one-sided, and mixed Gaussian IC. For the weak Gaussian IC, a new
outer bound on the capacity region is obtained that outperforms
previously known outer bounds. The sum capacity for a certain range
of channel parameters is derived. For this range, it is proved that
using Gaussian codebooks and treating interference as noise is
optimal. It is shown that when Gaussian codebooks are used, the full
Han-Kobayashi achievable rate region can be obtained by using the
naive Han-Kobayashi achievable scheme over three frequency bands
(equivalently, three subspaces). For the one-sided Gaussian IC, an
alternative proof for the Sato's outer bound is presented. We derive
the full Han-Kobayashi achievable rate region when Gaussian
codebooks are utilized. For the mixed Gaussian IC, a new outer bound
is obtained that outperforms previously known outer bounds. For this
case, the sum capacity for the entire range of channel parameters is
derived. It is proved that the full Han-Kobayashi achievable rate
region using Gaussian codebooks is equivalent to that of the
one-sided Gaussian IC for a particular range of channel parameters.
\end{abstract}


\section{Introduction}
\PARstart{O}{ne} of the fundamental problems in Information Theory,
originating from \cite{TWOWAY:SHANNON}, is the full characterization
of the capacity region of the interference channel (IC). The
simplest form of IC is the two-user case in which two transmitters
aim to convey independent messages to their corresponding receivers
through a common channel. Despite some special cases, such as very
strong and strong interference, where the exact capacity region has
been derived \cite{Very_Storng:Carleial,STRONG:SATO}, the
characterization of the capacity region for the general case is
still an open problem.

A limiting expression for the capacity region is obtained in
\cite{Ahlswede:Multi-way} (see also \cite{CSISZAR}). Unfortunately,
due to excessive computational complexity, this type of expression
does not result in a tractable approach to fully characterize the
capacity region. To show the weakness of the limiting expression,
Cheng and Verd\'{u} have shown that for the Gaussian Multiple Access
Channel (MAC), which can be considered as a special case of the
Gaussian IC, the limiting expression fails to fully characterize the
capacity region by relying only on Gaussian distributions
\cite{LE:CH-VER}. However, there is a point on the boundary of the
capacity region of the MAC that can be obtained directly from the
limiting expression. This point is achievable by using simple scheme
of Frequency/Time Division (FD/TD).

The computational complexity inherent to the limiting expression is
due to the fact that the corresponding encoding and decoding
strategies are of the simplest possible form. The encoding strategy
is based on mapping data to a codebook constructed from a unique
probability density and the decoding strategy is to treat the
interference as noise. In contrast, using more sophisticated
encoders and decoders may result in collapsing the limiting
expression into a single letter formula for the capacity region. As
an evidence, it is known that the joint typical decoder for the MAC
achieves the capacity region \cite{Cover-thomas}. Moreover, there
are some special cases, such as strong IC, where the exact
characterization of the capacity region has been derived
\cite{Very_Storng:Carleial,STRONG:SATO} where decoding the
interference is the key idea behind this result.

In their pioneering work, Han and Kobayashi (HK) proposed a coding
strategy in which the receivers are allowed to decode part of the
interference as well as their own data~\cite{IC:HK}. The HK
achievable region is still the best inner bound for the capacity
region. Specifically, in their scheme, the message of each user is
split into two independent parts: the common part and the private
part. The common part is encoded such that both users can decode it.
The private part, on the other hand, can be decoded only by the
intended receiver and the other receiver treats it as noise. In
summary, the HK achievable region is the intersection of the
capacity regions of two three-user MACs, projected on a
two-dimensional subspace.

The HK scheme can be directly applied to the Gaussian IC.
Nonetheless, there are two sources of difficulties in characterizing
the full HK achievable rate region. First, the optimal distributions
are unknown. Second, even if we confine the distributions to be
Gaussian, computation of the full HK region under Gaussian
distribution is still difficult due to numerous degrees of freedom
involved in the problem. The main reason behind this complexity is
the computation of the cardinality of the time-sharing parameter.

Recently, reference \cite{El-gamal-chong}, Chong {\em et al.} has
presented a simpler expression with less inequalities for the HK
achievable region. Since the cardinality of the time-sharing
parameter is directly related to the number of inequalities
appearing in the achievable rate region, the computational
complexity is decreased. However, finding the full HK achievable
region is still prohibitively complex.

Regarding outer bounds on the capacity region, there are three main
results known. The first one obtained by Sato \cite{ZIC:SATO} is
originally derived for the degraded Gaussian IC. Sato has shown that
the capacity region of the degraded Gaussian IC is outer bounded by
a certain degraded broadcast channel whose capacity region is fully
characterized. In \cite{IC:COSTA}, Costa has proved that the
capacity region of the degraded Gaussian broadcast channel is
equivalent to that of the one-sided weak Gaussian IC. Hence, Sato
outer bound can be used for the one-sided Gaussian IC as well.

The second outer bound obtained for the weak Gaussian IC is due to
Kramer \cite{BOUNDS:Kramer}. Kramer outer bound is based on the fact
that removing one of the interfering links enlarges the capacity
region. Therefore, the capacity region of the two-user Gaussian IC
is inside the intersection of the capacity regions of the underlying
one-sided Gaussian ICs. For the case of weak Gaussian IC, the
underlying one-sided IC is weak, for which the capacity region is
unknown. However, Kramer has used the outer bound obtained by Sato
to derive an outer bound for the weak Gaussian IC.

The third outer bound due to Etkin, Tse, and Wang (ETW) is based on
the Genie aided technique \cite{Etkin:IC}. A genie that provides
some extra information to the receivers can only enlarge the
capacity region. At first glance, it seems a clever genie must
provide some information about the interference to the receiver to
help in decoding the signal by removing the interference. In
contrast, the genie in the ETW scheme provides information about the
intended signal to the receiver. Remarkably, reference
\cite{Etkin:IC} shows that their proposed outer bound outperforms
Kramer bound for certain range of parameters. Moreover, using a
similar method, \cite{Etkin:IC} presents an outer bound for the
mixed Gaussian IC.

In this paper, by introducing the notion of admissible ICs, we
propose a new outer bounding technique for the two-user Gaussian IC.
The proposed technique relies on an extremal inequality recently
proved by Liu and Viswanath \cite{Liu:extremal}. We show that by
using this scheme, one can obtain tighter outer bounds for both weak
and mixed Gaussian ICs. More importantly, the sum capacity of the
Gaussian weak IC for a certain range of the channel parameters is
derived.

The rest of this paper is organized as follows. In Section II, we
present some basic definitions and review the HK achievable region
when Gaussian codebooks are used. We study the time-sharing and the
convexification methods as means to enlarge the basic HK achievable
region. We investigate conditions for which the two regions obtained
from time-sharing and concavification coincide. Finally, we consider
an optimization problem based on extremal inequality and compute its
optimal solution.

In Section III, the notion of an admissible IC is introduced. Some
classes of admissible ICs for the two-user Gaussian case is studied
and outer bounds on the capacity regions of these classes are
computed. We also obtain the sum capacity of a specific class of
admissible IC where it is shown that using Gaussian codebooks and
treating interference as noise is optimal.

In Section IV, we study the capacity region of the weak Gaussian IC.
We first derive the sum capacity of this channel for a certain range
of parameters where it is proved that users should treat the
interference as noise and transmit at their highest possible rates.
We then derive an outer bound on the capacity region which
outperforms the known results. We finally prove that the basic HK
achievable region results in the same enlarged region by using
either time-sharing or concavification. This reduces the complexity
of the characterization of the full HK achievable region when
Gaussian codebooks are used.

In Section V, we study the capacity region of the one-sided Gaussian
IC. We present a new proof for the Sato outer bound using the
extremal inequality. Then, we present methods to simplify the HK
achievable region such that the full region can be characterized.

In Section VI, we study the capacity region of the mixed Gaussian
IC. We first obtain the sum capacity of this channel and then derive
an outer bound which outperforms other known results. Finally, by
investigating the HK achievable region for different cases, we prove
that for a certain range of channel parameters, the full HK
achievable rate region using Gaussian codebooks is equivalent to
that of the one-sided IC. Finally, in Section VII, we conclude the
paper.

\subsection{Notations}
Throughout this paper, we use the following notations. Vectors are
represented by bold faced letters. Random variables, matrices, and
sets are denoted by capital letters where the difference is clear
from the context. $|A|$, $tr\{A\}$, and $A^t$ represent the
determinant, trace, and transpose of the square matrix $A$,
respectively. $I$ denotes the identity matrix. $\mathbb{N}$ and
$\Re$ are the sets of nonnegative integers and real numbers,
respectively. The union, intersection, and Minkowski sum of two sets
$U$ and $V$ are represented  by $U\cup V$, $U\cap V$, and $U+V$,
respectively. We use $\gamma(x)$ as an abbreviation for the function
$0.5\log_2(1+x)$.

\section{Preliminaries}

\subsection{The Two-user Interference Channel}

\begin{definition}[two-user IC]
A two-user discrete memoryless IC consists of two finite sets
$\mathscr{X}_1$ and $\mathscr{X}_2$ as input alphabets and two
finite sets $\mathscr{Y}_1$ and $\mathscr{Y}_2$ as the corresponding
output alphabets. The channel is governed by conditional probability
distributions $\omega{(y_1,y_2|x_1,x_2)}$ where
$(x_1,x_2)\in\mathscr{X}_1\times\mathscr{X}_2$ and
$(y_1,y_2)\in\mathscr{Y}_1\times\mathscr{Y}_2$.
\end{definition}

\begin{definition}[capacity region of the two-user IC]
A code ($2^{nR_{1}},2^{nR_{2}},n,\lambda_{1}^n,\lambda_{2}^n$) for
the two-user IC consists of the following components for User $i\in\{1,2\}$:

1) A uniform distributed message set
$\mathcal{M}_{i}\in[1,2,...,2^{nR_{i}}]$.

2) A codebook
$\mathcal{X}_{i}=\{\textbf{x}_{i}(1),\textbf{x}_{i}(2),...,\textbf{x}_{i}(2^{nR_{i}})\}$
where $\textbf{x}_{i}(\cdot)\in\mathscr{X}_i^n$.

3) An encoding function
$F_i:[1,2,...,2^{nR_{i}}]\rightarrow\mathcal{X}_{i}$.

4) A decoding function
$G_i:\textbf{y}_i\rightarrow[1,2,...,2^{nR_{i}}]$.

5) The average probability of error $\lambda_{i}^{n}=\mathbb{P}(G_i(\textbf{y}_i)\neq \mathcal{M}_i).$

A rate pair ($R_{1},R_{2}$) is achievable if there is a sequence of
codes ($2^{nR_{1}},2^{nR_{2}},n,\lambda_{1}^n,\lambda_{2}^n$) with
vanishing average error probabilities. The capacity region of the IC
is defined to be the supremum of the set of achievable rates.
\end{definition}

Let $\mathscr{C}_{IC}$ denote the capacity region of the two-user
IC. The limiting expression for $\mathscr{C}_{IC}$ can be stated as
\cite{CSISZAR}
\begin{eqnarray}\label{limiting expression}
\mathscr{C}_{IC}=\lim_{n\rightarrow\infty}{closure}\left(
\bigcup_{\mathbb{P}
(\textbf{X}_{1}^{n})\mathbb{P}(\textbf{X}_{2}^{n})}\left\{\left(R_{1},R_{2}\right)|
\begin{array}{c}
R_1\leq\frac{1}{n}\textbf{I}\left(\textbf{X}_{1}^{n},\textbf{Y}_{1}^{n}\right)\\
R_2\leq\frac{1}{n}\textbf{I}\left(\textbf{X}_{2}^{n},\textbf{Y}_{2}^{n}\right)
\end{array}
\right\}\right).
\end{eqnarray}

In this paper, we focus on the two-user Gaussian IC which can be
represented in standard form as \cite{IC:Carleial,IC:SASON}
\begin{equation}\label{Gaussian IC}
\begin{array}{rl}
y_{1}&=x_{1}+\sqrt{a}x_{2}+z_{1},\\
y_{2}&=\sqrt{b}x_{1}+x_{2}+z_{2},
\end{array}
\end{equation}
where $x_{i}$ and $y_{i}$ denote the input and output alphabets of
User $i\in\{1,2\}$, respectively, and $z_{1}\sim\mathcal{N}(0,1)$,
$z_{2}\sim\mathcal{N}(0,1)$ are standard Gaussian random variables.
Constants $a\geq0$ and $b\geq0$ represent the gains of the
interference links. Furthermore, Transmitter $i$, $i\in\{1,2\}$, is
subject to the power constraint $P_{i}$. Achievable rates and the
capacity region of the Gaussian IC can be defined in a similar
fashion as that of the general IC with the condition that the
codewords must satisfy their corresponding power constraints.
The capacity region of the two-user Gaussian IC is denoted by
$\mathscr{C}$. Clearly, $\mathscr{C}$ is a function of the
parameters $P_1$, $P_2$, $a$, and $b$. To emphasize this
relationship, we may write $\mathscr{C}$ as
$\mathscr{C}(P_1,P_2,a,b)$ as needed.

\begin{remark}\label{Remark: 1}
Since the capacity region of the general IC depends only on the
marginal distributions \cite{IC:SASON}, the ICs can be classified
into equivalent classes in which channels within a class have the
same capacity region. In particular, for the Gaussian IC given in
(\ref{Gaussian IC}), any choice of joint distributions for the pair
$(z_1,z_2)$ does not affect the capacity region as long as the
marginal distributions remain Gaussian with zero mean and unit
variance.
\end{remark}

\begin{figure}
\centering \includegraphics[scale=0.75]{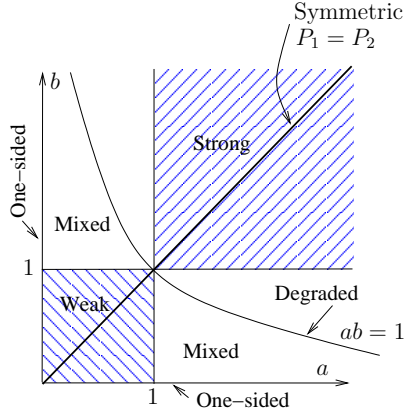}
\caption{Classes of the two-user ICs.}\label{fig sum6}
\end{figure}

Depending on the values of $a$ and $b$, the two-user Gaussian IC is
classified into weak, strong, mixed, one-sided, and degraded
Gaussian IC. In Figure \ref{fig sum6}, regions in $ab$-plane
together with their associated names are shown. Briefly, if $0<a<1$
and $0<b<1$, then the channel is called {\em weak Gaussian IC}. If
$1\leq a$ and $1\leq b$, then the channel is called {\em strong
Gaussian IC}. If either $a=0$ or $b=0$, the channel is called {\em
one-sided Gaussian IC}. If $ab=1$, then the channel is called {\em
degraded Gaussian IC}. If either $0<a<1$ and $1\leq b$, or $0<b<1$
and $1\leq a$, then the channel is called {\em mixed Gaussian IC}.
Finally, the {\em symmetric Gaussian IC} (used throughout the paper
for illustration purposes) corresponds to $a=b$ and $P_1=P_2$.

Among all classes shown in Figure~\ref{fig sum6}, the capacity
region of the strong Gaussian IC is fully characterized
\cite{STRONG:SATO,Very_Storng:Carleial}. In this case, the capacity
region can be stated as the collection of all rate pairs $(R_1,R_2)$
satisfying
\begin{equation}\nonumber
\begin{array}{rcl}
R_1&\leq &\gamma{(P_1)},\\
R_2&\leq &\gamma(P_2),\\
R_1+R_2&\leq & \min\left\{\gamma(P_1+aP_2),\gamma(bP_1+P_2)\right\}.
\end{array}
\end{equation}

\subsection{Support Functions}
Throughout this paper, we use the following facts from convex
analysis. There is a one to one correspondence between any closed
convex set and its support function \cite{BOYD:VANDENBERGHE}. The
support function of any set $D\in \Re^m$ is a function $\sigma_D
:\Re^m\rightarrow \Re$ defined as
\begin{equation}\label{support function}
\sigma_D (\mathbf{c})=\sup\{\mathbf{c}^t\mathbf{R}|\mathbf{R}\in D\}.
\end{equation}
Clearly, if the set $D$ is compact, then the sup is attained and can
be replaced by max. In this case, the solutions of (\ref{support
function}) correspond to the boundary points of $D$
\cite{BOYD:VANDENBERGHE}. The following relation is the dual of
(\ref{support function}) and holds when $D$ is closed and convex
\begin{equation}\label{dual representation}
D=\{\mathbf{R}|\mathbf{c}^t\mathbf{R}\leq\sigma_D (\mathbf{c}), \forall~\mathbf{c}\}.
\end{equation}
For any two closed convex sets $D$ and $D'$, $D\subseteq D'$, if and
only if $\sigma_D\leq\sigma_{D'}$.

\subsection{Han-Kobayashi Achievable Region}

The best inner bound for the two-user Gaussian IC is the full HK
achievable region denoted by $\mathscr{C}_{HK}$ \cite{IC:HK}.
Despite having a single letter formula, $\mathscr{C}_{HK}$ is not
fully characterized yet. In fact, finding the optimum distributions
achieving boundary points of $\mathscr{C}_{HK}$ is still an open
problem. We define $\mathscr{G}$ as a subset of $\mathscr{C}_{HK}$
where Gaussian distributions are used for codebook generation. Using
a shorter description of $\mathscr{C}_{HK}$ obtained in
\cite{El-gamal-chong}, $\mathscr{G}$ can be described as follows.

Let us first define $\mathscr{G}_0$ as the collection of all rate
pairs $(R_1,R_2)\in \Re_+^2$ satisfying
\begin{IEEEeqnarray}{rl}
\label{HK1}R_1 &\leq \psi_1=\gamma\left(\frac{P_1}{1+a\beta P_2}\right),\\
\label{HK2}R_2 &\leq  \psi_2=\gamma\left(\frac{P_2}{1+b\alpha P_1}\right),\\
\label{HK3}R_1+R_2 &\leq \psi_3=\min\left\{\psi_{31},\psi_{32},\psi_{33}\right\},\\
\label{HK4}2R_1+R_2 &\leq \psi_4= \gamma\left(\frac{P_1+a(1-\beta)P_2}{1+a\beta P_2}\right)+\gamma\left(\frac{\alpha P_1}{1+a\beta P_2}\right)+\gamma\left(\frac{\beta P_2+b(1-\alpha)P_1}{1+b\alpha P_1}\right),\\
\label{HK5}R_1+2R_2 &\leq \psi_5=\gamma\left(\frac{\beta P_2}{1+b\alpha P_1}\right)+\gamma\left(\frac{P_2+b(1-\alpha)P_1}{1+b\alpha P_1}\right)+\gamma\left(\frac{\alpha P_1+a(1-\beta)P_2}{1+a\beta P_2}\right),
\end{IEEEeqnarray}
for fixed $\alpha\in [0,1]$ and $\beta\in[0,1]$.\footnote{In the HK
scheme, two independent messages are encoded at each transmitter,
namely the {\em common message} and the {\em private message}.
$\alpha$ and $\beta$ are the parameters that determine the amount of
power allocated to the common and private messages for the two
users, i.e., $\alpha P_1$, $\beta P_2$ and $(1-\alpha)P_1$,
$(1-\beta)P_2$ of the total power is used for the transmission of
the private/common messages to the first/second users,
respectively.} $\psi_3$ is the minimum of $\psi_{31}$, $\psi_{32}$,
and $\psi_{33}$ defined as
\begin{IEEEeqnarray}{rl}
\label{HK6}\psi_{31}&= \gamma\left(\frac{P_1+a(1-\beta)P_2}{1+a\beta P_2}\right)+\gamma\left(\frac{\beta P_2}{1+b\alpha P_1}\right),\\
\label{HK7}\psi_{32}&= \gamma\left(\frac{\alpha P_1}{1+a\beta P_2}\right)+\gamma\left(\frac{P_2+b(1-\alpha)P_1}{1+b\alpha P_1}\right),\\
\label{HK8}\psi_{33}&= \gamma\left(\frac{\alpha
P_1+a(1-\beta)P_2}{1+a\beta P_2}\right)+\gamma\left(\frac{\beta
P_2+b(1-\alpha)P_1}{1+b\alpha P_1}\right).
\end{IEEEeqnarray}
$\mathscr{G}_0$ is a polytope and a function of four variables
$P_1$, $P_2$, $\alpha$, and $\beta$. To emphasize this relation, we
may write $\mathscr{G}_0(P_1,P_2,\alpha,\beta)$ as needed. It is
convenient to represent $\mathscr{G}_0$ in a matrix form as
$\mathscr{G}_0=\left\{\mathbf{R}|A\mathbf{R}\leq
\Psi(P_1,P_2,\alpha,\beta)\right\}$ where $\mathbf{R}=(R_1,R_2)^t$,
$\Psi=(\psi_1,\psi_2,\psi_3,\psi_4,\psi_5)^t$, and
\begin{equation}\nonumber
A=\left(
\begin{array}{ccccc}
1&0&1&2&1\\
0&1&1&1&2
\end{array}
\right)^t.
\end{equation}
Equivalently, $\mathscr{G}_0$ can be represented as the convex hull
of its extreme points, i.e.,
$\mathscr{G}_0(P_1,P_2,\alpha,\beta)=\text{conv}~\{r_1,r_2,\ldots,r_K\}$,
where it is assumed that $\mathscr{G}_0$ has $K$ extreme points. It
is easy to show that $K\leq 7$.

Now, $\mathscr{G}$ can be defined as a region obtained from
enlarging $\mathscr{G}_0$ by making use of the time-sharing
parameter, i.e., $\mathscr{G}$ is the collection of all rate pairs
$\mathbf{R}=(R_1,R_2)^t$ satisfying
\begin{IEEEeqnarray}{rl}
A\mathbf{R}&\leq \sum_{i=1}^q \lambda_i\Psi(P_{1i},P_{2i},\alpha_i,\beta_i),
\end{IEEEeqnarray}
where $q\in \mathbb{N}$ and
\begin{IEEEeqnarray}{rl}
\sum_{i=1}^q \lambda_iP_{1i}&\leq P_1,\\
\sum_{i=1}^q \lambda_iP_{2i}&\leq P_2,\\
\sum_{i=1}^q \lambda_i&=1,\\
\lambda_i\geq0,~(\alpha_i,\beta_i)&\in [0,1]^2;~\forall i\in\{1,\ldots,q\}.
\end{IEEEeqnarray}
It is easy to show that $\mathscr{G}$ is a closed, bounded and
convex region. In fact, the capacity region $\mathscr{C}$ which
contains $\mathscr{G}$ is inside the rectangle defined by
inequalities $R_1\leq \gamma(P_1)$ and $R_2\leq \gamma(P_2)$.
Moreover, $(0,0)$, $(\gamma(P_1),0)$, and $(0,\gamma(P_2))$ are
extreme points of both $\mathscr{C}$ and $\mathscr{G}$. Hence, to
characterize $\mathscr{G}$, we need to obtain all extreme points of
$\mathscr{G}$ that are in the interior of the first quadrant (the
same argument holds for $\mathscr{C}$). In other words, we need to
obtain $\sigma_{\mathscr{G}}(c_1,c_2)$, the support function of
$\mathscr{G}$, either when $1\leq c_1$ and $c_2=1$ or when $c_1=1$
and $1\leq c_2$.

We also define $\mathscr{G}_1$ and $\mathscr{G}_2$ obtained by
enlarging $\mathscr{G}_0$ in two different manners. $\mathscr{G}_1$
is defined as
\begin{equation}
\mathscr{G}_1(P_1,P_2)=\bigcup_{(\alpha,\beta)\in[0,1]^2}\mathscr{G}_0(P_1,P_2,\alpha,\beta).
\end{equation}
$\mathscr{G}_1$ is not necessarily a convex region. Hence, it can be
further enlarged by the convex hull operation. $\mathscr{G}_2$ is
defined as the collection of all rate pairs $\mathbf{R}=(R_1,R_2)^t$
satisfying
\begin{IEEEeqnarray}{rl}
\mathbf{R}=\sum_{i=1}^{q'} \lambda_i\mathbf{R}_i
\end{IEEEeqnarray}
where $q'\in \mathbb{N}$ and
\begin{IEEEeqnarray}{rl}
A\mathbf{R}_i&\leq \Psi(P_{1i},P_{2i},\alpha_i,\beta_i),\\
\sum_{i=1}^{q'} \lambda_iP_{1i}&\leq P_1,\\
\sum_{i=1}^{q'} \lambda_iP_{2i}&\leq P_2,\\
\sum_{i=1}^{q'} \lambda_i&=1,\\
\lambda_i\geq0,~(\alpha_i,\beta_i)&\in [0,1]^2;~\forall i\in\{1,\ldots,q'\}.
\end{IEEEeqnarray}
It is easy to show that $\mathscr{G}_2$ is a closed, bounded and
convex region. In fact, $\mathscr{G}_2$ is obtained by using the
simple method of TD/FD. To see this, let us divide the available
frequency band into $q'$ sub-bands where $\lambda_i$ represents the
length of the $i$'th band and $\sum_{i=1}^{q'}\lambda_i=1$. User 1
and 2 allocate $P_{1i}$ and $P_{2i}$ in the $i$'th sub-band,
respectively. Therefore, all rate pairs in
$\mathscr{G}_0(P_{1i},P_{2i},\alpha_i,\beta_i)$ are achievable in
the $i$'th sub-band for fixed $(\alpha_i,\beta_i)\in[0,1]^2$. Hence,
all rate pairs in $\sum_{i=1}^{q'}\lambda_i
\mathscr{G}_0(P_{1i},P_{2i},\alpha_i,\beta_i)$ are achievable
provided that $\sum_{i=1}^{q'} \lambda_iP_{1i}\leq P_1$ and
$\sum_{i=1}^{q'} \lambda_iP_{2i}\leq P_2$.

Clearly, the chain of inclusions
$\mathscr{G}_0\subseteq\mathscr{G}_1\subseteq \mathscr{G}_2
\subseteq \mathscr{G}\subseteq\mathscr{C}_{HK}\subseteq\mathscr{C}$
always holds.

\subsection{Concavification Versus Time-Sharing}
In this subsection, we follow two objectives. First, we aim at
providing some necessary conditions such that
$\mathscr{G}_2=\mathscr{G}$. Second, we bound $q$ and $q'$ which are
parameters involved in the descriptions of $\mathscr{G}$ and
$\mathscr{G}_2$, respectively. However, we derive the required
conditions for the more general case where there are $M$ users in
the system. To this end, assume an achievable scheme for an $M$-user
channel with the power constraint $\mathbf{P}=[P_1,P_2,\ldots,P_M]$
is given. The corresponding achievable region can be represented as
\begin{equation}\label{basic region}
D_0(\mathbf{P},\Theta)=\left\{\mathbf{R}|A\mathbf{R}\leq \Psi(\mathbf{P},\Theta)\right\},
\end{equation}
where $A$ is a $K\times M$ matrix and $\Theta\in[0,1]^M$. $D_0$ is a
polyhedron in general, but for the purpose of this paper, it
suffices to assume that it is a polytope. Since $D_0$ is a convex
region, the convex hull operation does not lead to a new enlarged
region. However, if the extreme points of the region are not a
concave function of $\mathbf{P}$, it is possible to enlarge $D_0$ by
using two different methods which are explained next. The first
method is based on using the time sharing parameter. Let us denote
the corresponding region as $D$ which can be written as
\begin{equation}\label{time-sharing}
D=\left\{\mathbf{R}|A\mathbf{R}\leq \sum_{i=1}^q \lambda_i\Psi(\mathbf{P}_i,\Theta_i),\sum_{i=1}^q \lambda_i\mathbf{P}_i\leq \mathbf{P},\sum_{i=1}^q \lambda_i=1,\lambda_i\geq0,\Theta_i\in[0,1]^M~\forall i\right\},
\end{equation}
where $q\in\mathbb{N}$.

In the second method, we use TD/FD to enlarge the achievable rate
region. This results in an achievable region $D_2$ represented as
\begin{equation}\label{concavification}
D_2=\left\{\mathbf{R}=\sum_{i=1}^{q'} \lambda_i\mathbf{R}_i|A\mathbf{R_i}\leq \Psi(\mathbf{P}_i,\Theta_i),\sum_{i=1}^{q'} \lambda_i\mathbf{P}_i\leq \mathbf{P},\sum_{i=1}^{q'} \lambda_i=1,\lambda_i\geq0,\Theta_i\in[0,1]^M~\forall i\right\},
\end{equation}
where $q'\in \mathbb{N}$. We refer to this method as
concavification. It can be readily shown that $D$ and $D_2$ are
closed and convex, and $D_2\subseteq D$. We are interested in
situations where the inverse inclusion holds.

The support function of $D_0$ is a function of $\mathbf{P}$, $\Theta$, and $\mathbf{c}$. Hence, we have
\begin{equation}\label{support function 1}
\sigma_{D_0} (\mathbf{c},\mathbf{P},\Theta)=\max\{\mathbf{c}^t\mathbf{R}|A\mathbf{R}\leq \Psi(\mathbf{P},\Theta)\}.
\end{equation}
For fixed $\mathbf{P}$ and $\Theta$, (\ref{support function 1}) is a
linear program. Using strong duality of linear programming, we
obtain
\begin{equation}\label{support function 2}
\sigma_{D_0}
(\mathbf{c},\mathbf{P},\Theta)=\min\{\mathbf{y}^t\Psi(\mathbf{P},\Theta)|A^t
\mathbf{y}=\mathbf{c},\mathbf{y}\geq 0 \}.
\end{equation}

In general, $\hat{\mathbf{y}}$, the minimizer of (\ref{support
function 2}), is a function of $\mathbf{P}$, $\Theta$, and
$\mathbf{c}$. We say $D_0$ possesses \emph{the unique minimizer
property} if $\hat{\mathbf{y}}$ merely depends on $\mathbf{c}$, for
all $\mathbf{c}$. In this case, we have
\begin{equation}\label{support function 3}
\sigma_{D_0} (\mathbf{c},\mathbf{P},\Theta)
=\hat{\mathbf{y}}^t(\mathbf{c})\Psi(\mathbf{P},\Theta),
\end{equation}
where $A^t \hat{\mathbf{y}}=\mathbf{c}$. This condition means that
for any $\mathbf{c}$ the extreme point of $D_0$ maximizing the
objective $\mathbf{c}^t\mathbf{R}$ is an extreme point obtained by
intersecting a set of specific hyperplanes. A necessary condition
for $D_0$ to possess the unique minimizer property is that each
inequality in describing $D_0$ is either redundant or active for all
$\mathbf{P}$ and $\Theta$.

\begin{theorem}\label{the unique minimizer property theorem}
If $D_0$ possesses the unique minimizer property, then $D=D_2$.
\end{theorem}
\begin{proof}
Since $D_2\subseteq D$ always holds, we need to show $D\subseteq
D_2$ which can be equivalently verified by showing $\sigma_{D}\leq
\sigma_{D_2}$. The support function of $D$ can be written as
\begin{equation}
\sigma_{D} (\mathbf{c},\mathbf{P})=\max
\left\{\mathbf{c}^t\mathbf{R}|\mathbf{R}\in D\right\}.
\end{equation}
By fixing $\mathbf{P}$, $\mathbf{P}_i$'s, $\Theta_i$'s, and
$\lambda_i$'s, the above maximization becomes a linear program.
Hence, relying on weak duality of linear programming, we obtain
\begin{equation}\label{support function 4}
\sigma_{D} (\mathbf{c},\mathbf{P})\leq
\min_{A^t\mathbf{y}=\mathbf{c},\mathbf{y}\geq0}
\mathbf{y}^t\sum_{i=1}^q \lambda_i\Psi(\mathbf{P}_i,\Theta_i).
\end{equation}
Clearly, $\hat{\mathbf{y}}(\mathbf{c})$, the solution of
(\ref{support function 2}), is a feasible point for (\ref{support
function 4}) and we have
\begin{equation}
\sigma_{D} (\mathbf{c},\mathbf{P})\leq
\hat{\mathbf{y}}^t(\mathbf{c})\sum_{i=1}^q
\lambda_i\Psi(\mathbf{P}_i,\Theta_i).
\end{equation}
Using (\ref{support function 3}), we obtain
\begin{equation}
\sigma_{D} (\mathbf{c},\mathbf{P})\leq \sum_{i=1}^q
\lambda_i\sigma_{D_0} (\mathbf{c},\mathbf{P}_i,\Theta_i).
\end{equation}
Let us assume $\hat{\mathbf{R}}_i$ is the maximizer of (\ref{support function 1}). In this case, we have
\begin{equation}
\sigma_{D} (\mathbf{c},\mathbf{P})\leq \sum_{i=1}^q \lambda_i \mathbf{c}^t\mathbf{\hat{R}}_i.
\end{equation}
Hence, we have
\begin{equation}
\sigma_{D} (\mathbf{c},\mathbf{P})\leq \mathbf{c}^t\sum_{i=1}^q \lambda_i \mathbf{\hat{R}}_i.
\end{equation}
By definition, $\sum_{i=1}^q \lambda_i \mathbf{\hat{R}}_i$ is a
point in $D_2$. Therefore, we conclude
\begin{equation}
\sigma_{D} (\mathbf{c},\mathbf{P})\leq \sigma_{D_2} (\mathbf{c},\mathbf{P}).
\end{equation}
This completes the proof.
\end{proof}

\begin{corollary}[Han \cite{Han:MAC-correlated-source}]
If $D_0$ is a polymatroid, then $D$=$D_2$.
\end{corollary}
\begin{proof}
It is easy to show that $D_0$ possesses the unique minimizer
property. In fact, for given $\mathbf{c}$, $\hat{\mathbf{y}}$ can be
obtained in a greedy fashion independent of $\mathbf{P}$ and
$\Theta$.
\end{proof}

In what follows, we upper bound $q$ and $q'$.
\begin{theorem}\label{caratheodory1}
The cardinality of the time sharing parameter $q$ in
(\ref{time-sharing}) is less than $M+K+1$, where $M$ and $K$ are the
dimensions of $\mathbf{P}$ and $\Psi(\mathbf{P})$, respectively.
Moreover, if $\Psi(\mathbf{P})$ is a continuous function of
$\mathbf{P}$, then $q\leq M+K$.
\end{theorem}
\begin{proof}
Let us define $E$ as
\begin{equation}
E=\left\{\sum_{i=1}^q \lambda_i\Psi(\mathbf{P}_i, \Theta_i)|\sum_{i=1}^q \lambda_i\mathbf{P}_i\leq \mathbf{P},\sum_{i=1}^q \lambda_i=1,\lambda_i\geq0, \Theta_i\in[0,1]^M~\forall i\right\}.
\end{equation}
In fact, $E$ is the collection of all possible bounds for $D$. To
prove $q\leq M+K+1$, we define another region $E_1$ as
\begin{equation}
E_1=\{(\mathbf{P}',\mathbf{S}')| 0\leq \mathbf{P}',\mathbf{S}'=\Psi(\mathbf{P}',\Theta'),\Theta'\in[0,1]^M\}.
\end{equation}
From the direct consequence of the Caratheodory's theorem
\cite{Rockafellar-wets}, the convex hull of $E_1$ denoted by
$\text{conv}~E_1$ can be obtained by convex combinations of no more
than $M+K+1$ points in $E_1$. Moreover, if
$\Psi(\mathbf{P}',\Theta')$ is continuous, then $M+K$ points are
sufficient due to the extension of the Caratheodory's theorem
\cite{Rockafellar-wets}. Now, we define the region $\hat{E}$ as
\begin{equation}
\hat{E}=\{\mathbf{S}'|(\mathbf{P}',\mathbf{S}')\in \text{conv}~E_1, \mathbf{P}'\leq \mathbf{P}\}.
\end{equation}
Clearly, $\hat{E}\subseteq E$. To show the other inclusion, let us
consider a point in $E$, say $S=\sum_{i=1}^q
\lambda_i\Psi(\mathbf{P}_i,\Theta_i)$. Since $(\mathbf{P}_i,
\Psi(\mathbf{P}_i,\Theta_i))$ is a point in $E_1$, $\sum_{i=1}^q
\lambda_i(\mathbf{P}_i,\Psi(\mathbf{P}_i,\Theta_i))$ belongs to
$\text{conv}~E_1$. Having $\sum_{i=1}^q
\lambda_i\mathbf{P}_i\leq\mathbf{P}$, we conclude $\sum_{i=1}^q
\lambda_i\Psi(\mathbf{P}_i,\Theta)\in \hat{E}$. Hence, $E\subseteq
\hat{E}$. This completes the proof.
\end{proof}

\begin{corollary}[Etkin, Parakh, and Tse \cite{Etkin-Spectrum}]
For the $M$-user Gaussian IC where users use Gaussian codebooks for
data transmission and treat the interference as noise, the
cardinality of the time sharing parameter is less than $2M$.
\end{corollary}
\begin{proof}
In this case, $D_0=\left\{\mathbf{R}|\mathbf{R}\leq
\Psi(\mathbf{P})\right\}$ where both $\mathbf{P}$ and
$\Psi(\mathbf{P})$ have dimension $M$ and $\Psi(\mathbf{P})$ is a
continuous function of $\mathbf{P}$. Applying Theorem
\ref{caratheodory1} yields the desired result.
\end{proof}

In the following theorem, we obtain an upper bound on $q'$.

\begin{theorem}\label{caratheodory}
To characterize boundary points of $D_2$, it suffices to set $q'\leq M+1$.
\end{theorem}
\begin{proof}
Let us assume $\hat{\mathbf{R}}$ is a boundary point of $D_2$.
Hence, there exists $\mathbf{c}$ such that
\begin{equation}\label{conv-1}
\sigma_{D_2} (\mathbf{c},\mathbf{P})=\max_{\mathbf{R}\in D_2}
\mathbf{c}^t\mathbf{R}=\mathbf{c}^t\hat{\mathbf{R}},
\end{equation}
where $\hat{\mathbf{R}}=\sum_{i=1}^{q'}\hat{\lambda}_i
\hat{\mathbf{R}}_i$ and the optimum is achieved for the set of
parameters $\hat{\Theta}_i$, $\hat{\lambda}_i$, and
$\hat{\mathbf{P}}_i$. The optimization problem in (\ref{conv-1}) can
be written as
\begin{IEEEeqnarray}{rl}\label{conv-2}
\sigma_{D_2} (\mathbf{c},\mathbf{P})=&\max ~~~\sum_{i=1}^{q'} \lambda_i g(\mathbf{c},\mathbf{P}_i)\\
&\text{subject to:}~\sum_{i=1}^{q'} \lambda_i=1\nonumber,~\sum_{i=1}^{q'} \lambda_i\mathbf{P}_i\leq \mathbf{P},\\
&~~~~~~~~~~~~~~0\leq\lambda_i,0\leq \mathbf{P}_i, ~\forall
i\in\{1,2,\ldots,q'\},\nonumber
\end{IEEEeqnarray}
where $g(\mathbf{c},\mathbf{P})$ is defined as
\begin{IEEEeqnarray}{rl}\label{conv-3}
g(\mathbf{c},\mathbf{P})=&\max \mathbf{c}^t\mathbf{R}\\
&\text{subject to:}~A\mathbf{R}\leq \Psi(\mathbf{P},\Theta),~ 0\leq
\Theta\leq 1,\nonumber
\end{IEEEeqnarray}

In fact, $\sigma_{D_2} (\mathbf{c},\mathbf{P})$ in (\ref{conv-2})
can be viewed as the result of the concavification of
$g(\mathbf{c},\mathbf{P})$ \cite{Rockafellar-wets}. Hence, using
Theorem 2.16 in \cite{Rockafellar-wets}, we conclude that $q'\leq
M+1$.
\end{proof}

Remarkable point about Theorem \ref{caratheodory} is that the upper
bound on $q'$ is independent of the number of inequalities involved
in the description of the achievable rate region.

\begin{corollary}
For the $M$-user Gaussian IC where users use Gaussian codebooks and
treat the interference as noise, we have $D_2=D$ and $q=q'=M+1$.
\end{corollary}

\subsection{Extremal Inequality}
In \cite{Liu:extremal}, the following optimization problem is studied:
\begin{equation}\label{extremal-general}
W=\max_{Q_\mathbf{X}\leq S} h(\mathbf{X}+\mathbf{Z}_1)-\mu h(\mathbf{X}+\mathbf{Z}_2),
\end{equation}
where $\mathbf{Z}_1$ and $\mathbf{Z}_2$ are $n$-dimensional Gaussian
random vectors with the strictly positive definite covariance
matrices $Q_{\mathbf{Z}_1}$ and $Q_{\mathbf{Z}_2}$, respectively.
The optimization is over all random vectors $\mathbf{X}$ independent
of $\mathbf{Z}_1$  and $\mathbf{Z}_2$. $\mathbf{X}$ is also subject
to the covariance matrix constraint $Q_{\mathbf{X}}\leq S$, where
$S$ is a positive definite matrix. In \cite{Liu:extremal}, it is
shown that for all $\mu\geq 1$, this optimization problem has a
Gaussian optimal solution for all positive definite matrices
$Q_{\mathbf{Z}_1}$ and $Q_{\mathbf{Z}_2}$. However, for $0\leq\mu<
1$ this optimization problem has a Gaussian optimal solution
provided $Q_{\mathbf{Z}_1}\leq Q_{\mathbf{Z}_2}$, i.e.,
$Q_{\mathbf{Z}_2}-Q_{\mathbf{Z}_1}$ is a positive semi-definite
matrix. It is worth noting that for $\mu=1$ this problem when
$Q_{\mathbf{Z}_1}\leq Q_{\mathbf{Z}_2}$ is studied under the name of
the worse additive noise
\cite{Diggavi:worst-noise,Ihara:non-Gaussian}.

In this paper, we consider a special case of
(\ref{extremal-general}) where $\mathbf{Z}_1$ and $\mathbf{Z}_2$
have the covariance matrices $N_1I$ and $N_2I$, respectively, and
the trace constraint is considered, i.e.,
\begin{equation}\label{extremal}
W=\max_{tr\{Q_\mathbf{X}\}\leq nP} h(\mathbf{X}+\mathbf{Z}_1)-\mu h(\mathbf{X}+\mathbf{Z}_2).
\end{equation}
In the following lemma, we provide the optimal solution for the above optimization problem when $N_1\leq N_2$.

\begin{lemma}\label{lem Extremal-1}
If $N_1\leq N_2$, the optimal solution of (\ref{extremal}) is iid
Gaussian for all $0\leq\mu$ and we have
\begin{enumerate}
\item For $0\leq \mu \leq \frac{N_2+P}{N_1+P}$, the optimum covariance matrix is $PI$ and the optimum solution is
    \begin{equation}
    W=\frac{n}{2}\log\left[(2\pi e)(P+N_1)\right]-\frac{\mu n}{2}\log\left[(2\pi e)(P+N_2)\right].
    \end{equation}
\item For $\frac{N_2+P}{N_1+P}< \mu \leq \frac{N_2}{N_1}$, the optimum covariance matrix is $\frac{N_2-\mu N_1}{\mu-1}I$ and the optimum solution is
    \begin{equation}
    W=\frac{n}{2}\log\left[(2\pi e)\frac{N_2-N_1}{\mu-1}\right]-\frac{\mu n}{2}\log\left[\frac{\mu(2\pi
    e)(N_2-N_1)}{\mu-1}\right].
    \end{equation}
\item For $\frac{N_2}{N_1}< \mu$, the optimum covariance matrix is $0$ and the optimum solution is
    \begin{equation}
    W=\frac{n}{2}\log(2\pi e N_1)-\frac{\mu n}{2}\log(2\pi eN_2).
    \end{equation}
\end{enumerate}
\end{lemma}

\begin{proof}
From the general result for (\ref{extremal-general}), we know that
the optimum input distribution is Gaussian. Hence, we need to solve
the following maximization problem:
\begin{IEEEeqnarray}{rl}
W=&\max \frac{1}{2}\log\left((2\pi e)^n|Q_\mathbf{X}+N_1I|\right)
-\frac{\mu}{2}\log\left((2\pi e)^n|Q_\mathbf{X}+N_2I|\right)\label{extremal-1}\\
&\text{subject to:}~0\leq Q_\mathbf{X},~tr\{Q_\mathbf{X}\}\leq
nP.\nonumber
\end{IEEEeqnarray}
Since $Q_\mathbf{X}$ is a positive semi-definite matrix, it can be
decomposed as $Q_\mathbf{X}=U\Lambda U^t$, where $\Lambda$ is a
diagonal matrix with nonnegative entries and $U$ is a unitary
matrix, i.e., $UU^t=I$. Substituting $Q_\mathbf{X}=U\Lambda U^t$ in
(\ref{extremal-1}) and using the identities $tr\{AB\}=tr\{BA\}$ and
$|AB+I|=|BA+I|$, we obtain
\begin{IEEEeqnarray}{rl}
W=&\max \frac{1}{2}\log\left((2\pi e)^n|\Lambda+N_1I|\right)
-\frac{\mu}{2}\log\left((2\pi e)^n|\Lambda+N_2I|\right)\\
&\text{subject to:}~0\leq \Lambda,~tr\{\Lambda\}\leq nP.\nonumber
\end{IEEEeqnarray}
This optimization problem can be simplified as
\begin{IEEEeqnarray}{rl}
W=&\max \frac{n}{2}\sum_{i=1}^n \left[\log(2\pi e)(\lambda_i+N_1)-\mu\log(2\pi e)(\lambda_i+N_2)\right]\\
&\text{subject to:}~0\leq \lambda_i~\forall i,~\sum_{i=1}^n
\lambda_i\leq nP.\nonumber
\end{IEEEeqnarray}
By introducing Lagrange multipliers $\psi$ and
$\Phi=\{\phi_1,\phi_2,\ldots,\phi_n\}$, we obtain
\begin{equation}\label{extremal-2}
L(\Lambda,\psi,\Phi)=\max \frac{n}{2}\sum_{i=1}^n \left[\log(2\pi e)(\lambda_i+N_1)-\mu\log(2\pi e)(\lambda_i+N_2)\right]+ \psi\left(nP-\sum_{i=1}^n\lambda_i\right)+\sum_{i=1}^n\phi_i\lambda_i.
\end{equation}

The first order KKT necessary conditions for the optimum solution of
(\ref{extremal-2}) can be written as
\begin{IEEEeqnarray}{rl}
\frac{1}{\lambda_i+N_1}-\frac{\mu}{\lambda_i+N_2}-\psi+\phi_i=&0,~\forall i\in\{1,2,\ldots,n\},\\
\psi\left(nP-\sum_{i=1}^n\lambda_i\right)=&0,\\
\phi_i\lambda_i=&0, ~\forall i\in\{1,2,\ldots,n\}.
\end{IEEEeqnarray}

It is easy to show that when $N_1\leq N_2$, $\lambda=\lambda_1=\ldots=\lambda_n$ and the only solution for $\lambda$ is
\begin{equation}
\lambda=\left\{
\begin{array}{llrl}
P,&~ \text{if}~& 0&\leq \mu \leq \frac{N_2+P}{N_1+P}\\
\frac{N_2-\mu N_1}{\mu-1},&~ \text{if}~& \frac{N_2+P}{N_1+P}&< \mu \leq \frac{N_2}{N_1}\\
0,&~ \text{if}~& \frac{N_2}{N_1}&< \mu
\end{array}
\right.
\end{equation}
Substituting $\lambda$ into the objective function gives the desired
result.
\end{proof}

\begin{figure}
\centering \includegraphics[scale=0.75]{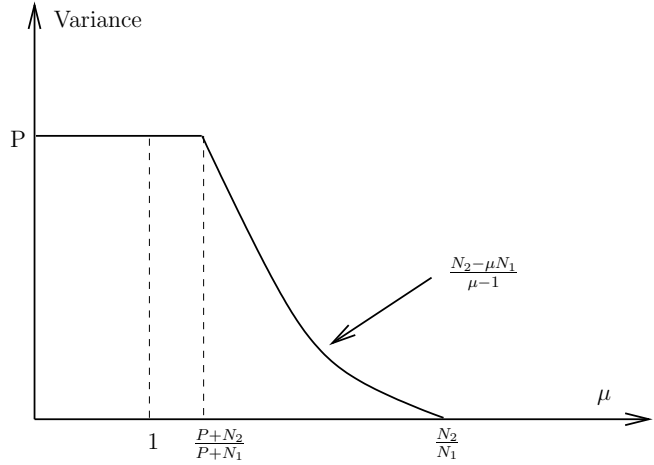}
\caption{Optimum variance versus $\mu$.}\label{fig sum5}
\end{figure}

In Figure \ref{fig sum5}, the optimum variance as a function of
$\mu$ is plotted. This figure shows that for any value of $\mu\leq
\frac{P+N_2}{P+N_1}$, we need to use the maximum power to optimize
the objective function, whereas for $\mu> \frac{P+N_2}{P+N_1}$, we
use less power than what is permissible.

\begin{lemma}
If $N_1> N_2$, the optimal solution of (\ref{extremal}) is iid
Gaussian for all $1\leq\mu$. In this case, the optimum variance is
$0$ and the optimum $W$ is
\begin{equation}
W=\frac{n}{2}\log(2\pi e N_1)-\frac{\mu n}{2}\log(2\pi eN_2).
\end{equation}
\end{lemma}
\begin{proof}
The proof is similar to that of Lemma \ref{lem Extremal-1} and we omit it here.
\end{proof}

\begin{corollary}\label{lem Extremal}
For $\mu=1$, the optimal solution of (\ref{extremal}) is iid
Gaussian and the optimum $W$ is
\begin{equation}
W=\left\{
\begin{array}{ll}
\frac{n}{2}\log\left(\frac{P+N_1}{P+N_{2}}\right), & \text{if}~ N_1\leq N_2\\
\frac{n}{2}\log\left(\frac{N_1}{N_{2}}\right), & \text{if}~ N_1>
N_2.
\end{array}
\right.
\end{equation}
\end{corollary}

We frequently apply the following optimization problem in the rest
of the paper:
\begin{equation}\label{fh-function}
f_h(P,N_1,N_2,a,\mu)=\max_{tr\{Q_\mathbf{X}\}\leq nP}
h(\mathbf{X}+\mathbf{Z}_1)-\mu h(\sqrt{a}\mathbf{X}+\mathbf{Z}_2),
\end{equation}
where $N_1\leq N_2/a$. Using the identity $h(A\mathbf{X})=\log(|A|)+h(\mathbf{X})$, (\ref{fh-function}) can be written as
\begin{equation}
f_h(P,N_1,N_2,a,\mu)=\frac{n}{2}\log a+\max_{tr\{Q_\mathbf{X}\}\leq
nP} h(\mathbf{X}+\mathbf{Z}_1)-\mu
h(\mathbf{X}+\frac{\mathbf{Z}_2}{\sqrt{a}}).
\end{equation}
Now, Lemma \ref{lem Extremal-1} can be applied to obtain
\begin{equation}\label{extremal-function}
f_h(P,N_1,N_2,a,\mu)=\left\{
\begin{array}{ll}
\frac{1}{2}\log\left[(2\pi e)(P+N_1)\right]-\frac{\mu}{2}\log\left[(2\pi e)(aP+N_2)\right]&~\text{if}~ 0\leq \mu \leq \frac{P+N_2/a}{P+N_1}\\
\frac{1}{2}\log\left[(2\pi e)\frac{N_2/a-N_1}{\mu-1}\right]- \frac{\mu}{2}\log\left[\frac{a\mu(2\pi e)(N_2/a-N_1)}{\mu-1}\right]&~\text{if}~ \frac{P+N_2/a}{P+N_1}< \mu \leq \frac{N_2}{aN_1}\\
\frac{1}{2}\log(2\pi e N_1)-\frac{\mu}{2}\log(2\pi eN_2)&~\text{if}~\frac{N_2}{aN_1}<\mu
\end{array}
\right.
\end{equation}

\section{Admissible Channels}
In this section, we aim at building ICs whose capacity regions
contain the capacity region of the two-user Gaussian IC, i.e.,
$\mathscr{C}$. Since we ultimately use these to outer bound
$\mathscr{C}$, these ICs need to have a tractable expression (or a
tractable outer bound) for their capacity regions.

Let us consider an IC with the same input letters as that of
$\mathscr{C}$ and the output letters $\tilde{y}_1$ and $\tilde{y}_2$
for Users 1 and 2, respectively. The capacity region of this
channel, say $\mathscr{C}'$, contains $\mathscr{C}$ if
\begin{IEEEeqnarray}{rl}
I(x_1^n;y_1^n)\leq &I(x_1^n;\tilde{y}_{1}^n)\label{condition1},\\
I(x_2^n;y_2^n)\leq &I(x_2^n;\tilde{y}_{2}^n)\label{condition2},
\end{IEEEeqnarray}
for all $p(x_1^n)p(x_2^n)$ and for all $n\in \mathbb{N}$.

One way to satisfy (\ref{condition1}) and (\ref{condition2}) is to
provide some extra information to either one or to both receivers.
This technique is known as {\em Genie aided outer bounding}. In
\cite{BOUNDS:Kramer}, Kramer has used such a genie to provide some
extra information to both receivers such that they can decode both
users' messages. Since the capacity region of this new interference
channel is equivalent to that of the {\em Compound Multiple Access
Channel} whose capacity region is known, reference
\cite{BOUNDS:Kramer} obtains an outer bound on the capacity region.
To obtain a tighter outer bound, reference \cite{BOUNDS:Kramer}
further uses the fact that if a genie provides the exact information
about the interfering signal to one of the receivers, then the new
channel becomes the one-sided Gaussian IC. Although the capacity
region of the one-sided Gaussian IC is unknown for all ranges of
parameters, there exists an outer bound for it due to Sato and Costa
\cite{Sato:BC-outer,IC:COSTA} that can be applied to the original
channel. In \cite{Etkin:IC}, Etkin {\em et al.} use a different
genie that provides some extra information about the intended
signal. Even though at first glance their proposed method appears to
be far from achieving a tight bound, remarkably they show that the
corresponding bound is tighter than the one due to Kramer for
certain ranges of parameters.

Next, we introduce the notion of admissible channels to satisfy
(\ref{condition1}) and (\ref{condition2}).

\begin{figure}
\centering \includegraphics[scale=0.75]{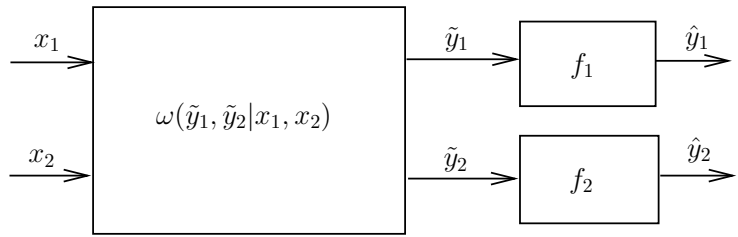}
\caption{An admissible channel. $f_1$ and $f_2$ are deterministic
functions.}\label{admissible channel}
\end{figure}

\begin{definition}[Admissible Channel]
An IC $\mathscr{C}'$ with input letter $x_i$ and output letter
$\tilde{y}_i$ for User $i\in\{1,2\}$ is an admissible channel if
there exist two deterministic functions
$\hat{y}_1^n=f_1(\tilde{y}_{1}^n)$ and
$\hat{y}_2^n=f_2(\tilde{y}_{2}^n)$ such that
\begin{IEEEeqnarray}{rl}
I(x_1^n;y_1^n)\leq &I(x_1^n;\hat{y}_{1}^n)\label{condition3},\\
I(x_2^n;y_2^n)\leq &I(x_2^n;\hat{y}_{2}^n)\label{condition4}
\end{IEEEeqnarray}
hold for all $p(x_1^n)p(x_2^n)$ and for all $n\in \mathbb{N}$.
$\mathscr{E}$ denotes the collection of all admissible channels (see
Figure \ref{admissible channel}).
\end{definition}

\begin{remark}
Genie aided channels are among admissible channels. To see this, let
us assume a genie provides $s_1$ and $s_2$ as side information for
User 1 and 2, respectively. In this case, $\tilde{y}_{i}=(y_i,s_i)$
for $i\in\{1,2\}$. By choosing $f_i(y_i,s_i)=y_i$, we observe that
$\hat{y}_{i}=y_i$, and hence, (\ref{condition3}) and
(\ref{condition4}) trivially hold.
\end{remark}

To obtain the tightest outer bound, we need to find the intersection
of the capacity regions of all admissible channels. Nonetheless, it
may happen that finding the capacity region of an admissible channel
is as hard as that of the original one (in fact, based on the
definition, the channel itself is one of its admissible channels).
Hence, we need to find classes of admissible channels, say
$\mathscr{F}$, which possess two important properties. First, their
capacity regions are close to $\mathscr{C}$. Second, either their
exact capacity regions are computable or there exist good outer
bounds for them. Since $\mathscr{F}\subseteq \mathscr{E}$, we have
\begin{equation}
\mathscr{C}\subseteq \bigcap_{\mathscr{F}} \mathscr{C}'.
\end{equation}
Recall that there is a one to one correspondence between a closed
convex set and its support function. Since $\mathscr{C}$ is closed
and convex, there is a one to one correspondence between
$\mathscr{C}$ and $\sigma_{\mathscr{C}}$. In fact, boundary points
of $\mathscr{C}$ correspond to the solutions of the following
optimization problem
\begin{equation}\label{raletion1}
\sigma_{\mathscr{C}}(c_1,c_2)=\max_{(R_1,R_2)\in \mathscr{C}}
c_1R_1+c_2R_2.
\end{equation}
Since we are interested in the boundary points excluding the $R_1$
and $R_2$ axes, it suffices to consider $0\leq c_1$ and $0\leq c_2$
where $c_1+c_2=1$.

Since $\mathscr{C}\subseteq \mathscr{C}'$, we have
\begin{equation}
\sigma_{\mathscr{C}}(c_1,c_2)\leq \sigma_{\mathscr{C}'}(c_1,c_2).
\end{equation}
Taking the minimum of the right hand side, we obtain
\begin{equation}
\sigma_{\mathscr{C}}(c_1,c_2)\leq \min_{\mathscr{C}'\in \mathscr{F}} \sigma_{\mathscr{C}'}(c_1,c_2),
\end{equation}
which can be written as
\begin{equation}
\sigma_{\mathscr{C}}(c_1,c_2)\leq \min_{\mathscr{C}'\in \mathscr{F}} \max_{(R_1,R_2)\in \mathscr{C}'} c_1 R_1+c_2R_2.
\end{equation}

For convenience, we use the following two optimization problems
\begin{equation}\label{raletion2}
\sigma_{\mathscr{C}}(\mu,1)=\max_{(R_1,R_2)\in \mathscr{C}} \mu
R_1+R_2,
\end{equation}
\begin{equation}\label{raletion3}
\sigma_{\mathscr{C}}(1,\mu)=\max_{(R_1,R_2)\in \mathscr{C}} R_1+\mu R_2,
\end{equation}
where $1\leq \mu$. It is easy to show that the solutions of
(\ref{raletion2}) and (\ref{raletion3}) correspond to the boundary
points of the capacity region.

In the rest of this section, we introduce classes of admissible
channels and obtain upper bounds on $\sigma_{\mathscr{C}'}(\mu,1)$
and $\sigma_{\mathscr{C}'}(1,\mu)$.

\subsection{Classes of Admissible Channels}

\subsubsection{Class A1}
\begin{figure}
\centering \includegraphics[scale=0.65]{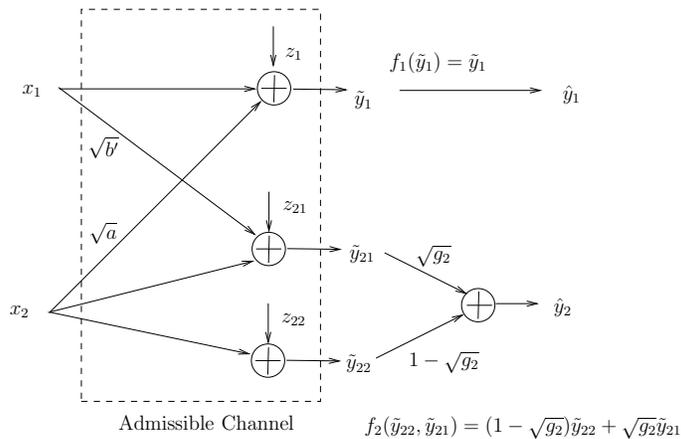}
\caption{Class A1 admissible channels.}\label{classa1 admissible channels}
\end{figure}

This class is designed to obtain an upper bound on
$\sigma_{\mathscr{C}}(\mu,1)$. Therefore, we need to find a tight
upper bound on $\sigma_{\mathscr{C}'}(\mu,1)$. A member of this
class is a channel in which User 1 has one transmit and one receive
antenna whereas User 2 has one transmit antenna and two receive
antennas (see Figure \ref{classa1 admissible channels}). The channel
model can be written as
\begin{equation}\label{class A1}
\begin{array}{rl}
\tilde{y}_{1}=&x_{1}+\sqrt{a}x_{2}+z_{1},\\
\tilde{y}_{21}=&x_{2}+\sqrt{b'}x_{1}+z_{21},\\
\tilde{y}_{22}=&x_{2}+z_{22},
\end{array}
\end{equation}
where $\tilde{y}_{1}$ is the signal at the first receiver,
$\tilde{y}_{21}$ and $\tilde{y}_{22}$ are the signals at the second
receiver, $z_{1}$ is additive Gaussian noise with unit variance,
$z_{21}$ and $z_{22}$ are additive Gaussian noise with variances
$N_{21}$ and $N_{22}$, respectively. Transmitters 1 and 2 are
subject to the power constraints of $P_1$ and $P_2$, respectively.

To investigate admissibility conditions in (\ref{condition3}) and
(\ref{condition4}), we introduce two deterministic functions $f_1$
and $f_2$ as follows (see Figure \ref{classa1 admissible channels})
\begin{IEEEeqnarray}{rl}
f_1(\tilde{y}_{1}^n)&=\tilde{y}_{1}^n,\\
f_2(\tilde{y}_{22}^n,\tilde{y}_{21}^n)&= (1-\sqrt{g_2})\tilde{y}_{22}^n+\sqrt{g_2}\tilde{y}_{21}^n,
\end{IEEEeqnarray}
where $0\leq g_2$. For $g_2=0$, the channel can be converted to the
one-sided Gaussian IC by letting $N_{21}\rightarrow \infty$ and
$N_{22}=1$. Hence, Class A1 contains the one-sided Gaussian IC
obtained by removing the link between Transmitter 1 and Receiver 2.
Using $f_1$ and $f_2$, we obtain
\begin{IEEEeqnarray}{rl}
\hat{y}_1^n=&x_1^n+\sqrt{a}x_2^n+z_{1}^n,\\
\hat{y}_2^n=&\sqrt{b'g_2}x_1^n+x_2^n+(1-\sqrt{g_2})z_{22}^n+\sqrt{g_2}z_{21}^n.
\end{IEEEeqnarray}
Hence, this channel is admissible if the corresponding parameters
satisfy
\begin{equation}\label{class A1 conditions}
\begin{array}{rl}
b'g_2&=b,\\
(1-\sqrt{g_2})^2N_{22}+g_2N_{21}&=1.
\end{array}
\end{equation}
We further add the following constraints to the conditions of the
channels in Class A1:
\begin{equation}\label{class A1 conditions1}
\begin{array}{rl}
b'&\leq N_{21},\\
aN_{22}&\leq 1.
\end{array}
\end{equation}
Although these additional conditions reduce the number of admissible
channels within the class, they are needed to get a closed form
formula for an upper bound on $\sigma_{\mathscr{C}'}(\mu,1)$. In the
following lemma, we obtain the required upper bound.

\begin{lemma}\label{class A1 lemma}
For the channels modeled by (\ref{class A1}) and satisfying
(\ref{class A1 conditions1}), we have
\begin{IEEEeqnarray}{rl}
\sigma_{\mathscr{C}'}(\mu,1)\leq&\min \frac{\mu_1}{2}\log\left[2\pi
e(P_1+aP_2+1)\right]-\frac{\mu_2}{2}\log(2\pi e)
+\frac{1}{2}\log\left( \frac{N_{21}}{N_{22}}+\frac{b'P_1}{N_{22}}+\frac{P_2 }{P_2+N_{22}}\right)\label{class A1 solution}\\
&~~~~~~~~~~~~+\mu_2 f_h\left(P_1,1,N_{21},b',\frac{1}{\mu_2}\right) +f_h(P_2,N_{22},1,a,\mu_1)\nonumber\\
&\text{subject to:}~\mu_1+\mu_2=\mu,~\mu_1,\mu_2\geq 0.\nonumber
\end{IEEEeqnarray}

\end{lemma}
\begin{proof}
Let us assume $R_1$ and $R_2$ are achievable rates for User 1 and 2,
respectively. Furthermore, we split $\mu$ into $\mu_1\geq 0$ and
$\mu_2\geq0$ such that $\mu=\mu_1+\mu_2$. Using Fano's inequality,
we obtain
\begin{IEEEeqnarray}{rl}
n(\mu R_1+R_2)\stackrel{}{\leq}& \mu I(x_1^n;\tilde{y}_1^n)+ I(x_2^n;\tilde{y}_{22}^n,\tilde{y}_{21}^n)+n\epsilon_n\nonumber\\
\stackrel{}{=}& \mu_1 I(x_1^n;\tilde{y}_1^n)+\mu_2 I(x_1^n;\tilde{y}_1^n)+ I(x_2^n;\tilde{y}_{22}^n,\tilde{y}_{21}^n)+n\epsilon_n\nonumber\\
\stackrel{(a)}{\leq}& \mu_1 I(x_1^n;\tilde{y}_1^n)+\mu_2 I(x_1^n;\tilde{y}_1^n|x_2^n)+ I(x_2^n;\tilde{y}_{22}^n,\tilde{y}_{21}^n)+n\epsilon_n\nonumber\\
\stackrel{}{=}& \mu_1 I(x_1^n;\tilde{y}_1^n)+\mu_2 I(x_1^n;\tilde{y}_1^n|x_2^n)+I(x_2^n;\tilde{y}_{21}^n|\tilde{y}_{22}^n) +I(x_2^n;\tilde{y}_{22}^n)+n\epsilon_n\nonumber\\
\stackrel{}{=}& \mu_1 h(\tilde{y}_1^n)-\mu_1 h(\tilde{y}_1^n|x_1^n)+\mu_2 h(\tilde{y}_1^n|x_2^n)-\mu_2 h(\tilde{y}_1^n|x_1^n,x_2^n)\nonumber\\
&+h(\tilde{y}_{21}^n|\tilde{y}_{22}^n)-h(\tilde{y}_{21}^n|x_2^n,\tilde{y}_{22}^n) +h(\tilde{y}_{22}^n)-h(\tilde{y}_{22}^n|x_2^n)+n\epsilon_n\nonumber\\
\stackrel{}{=}& \big[\mu_1 h(\tilde{y}_1^n)-\mu_2 h(\tilde{y}_1^n|x_1^n,x_2^n)\big]+\big[\mu_2 h(\tilde{y}_1^n|x_2^n)-h(\tilde{y}_{21}^n|x_2^n,\tilde{y}_{22}^n)\big]\nonumber\\
&+\big[h(\tilde{y}_{21}^n|\tilde{y}_{22}^n)-h(\tilde{y}_{22}^n|x_2^n)\big] +\big[h(\tilde{y}_{22}^n)-\mu_1 h(\tilde{y}_1^n|x_1^n)\big]+n\epsilon_n,\label{class A1 fano}
\end{IEEEeqnarray}
where (a) follows from the fact that $x_1^n$ and $x_2^n$ are
independent. Now, we separately upper bound the terms within each
bracket in (\ref{class A1 fano}).

To maximize the terms within the first bracket, we use the fact that
Gaussian distribution maximizes the differential entropy subject to
a constraint on the covariance matrix. Hence, we have
\begin{IEEEeqnarray}{rl}
\mu_1 h(\tilde{y}_1^n)-\mu_2 h(\tilde{y}_1^n|x_1^n,x_2^n) &=\mu_1 h(x_1^n+\sqrt{a}x_2^n+z_1^n)-\mu_2 h(z_1^n)\nonumber\\
&\leq \frac{\mu_1 n}{2}\log\left[2\pi e(P_1+aP_2+1)\right]-\frac{\mu_2 n}{2}\log(2\pi e).
\end{IEEEeqnarray}

Since $b'\leq N_{21}$, we can make use of Lemma \ref{lem Extremal-1}
to upper bound the second bracket. In this case, we have
\begin{IEEEeqnarray}{rl}
\mu_2 h(\tilde{y}_{1}^n|x_2^n)-h(\tilde{y}_{21}^n|x_2^n,\tilde{y}_{22}^n)&= \mu_2 \left(h(x_1^n+z_{1}^n)-\frac{1}{\mu_2}h(\sqrt{b'}x_1^n+z_{21}^n)\right)\nonumber\\
&\leq \mu_2 nf_h\left(P_1,1,N_{21},b',\frac{1}{\mu_2}\right),
\end{IEEEeqnarray}
where $f_h$ is defined in (\ref{extremal-function}).

We upper bound the terms within the third bracket as follows
\cite{Etkin:IC}:
\begin{IEEEeqnarray}{rl}
h(\tilde{y}_{21}^n|\tilde{y}_{22}^n)-h(\tilde{y}_{22}^n|x_2^n)\stackrel{(a)}{\leq}& \sum_{i=1}^n h(\tilde{y}_{21}[i]|\tilde{y}_{22}[i])-h(z_{22}^n)\nonumber\\
\stackrel{(b)}{\leq}& \sum_{i=1}^n\frac{1}{2}\log\left[2\pi e\left( N_{21}+b'P_1[i]+\frac{P_2[i]N_{22}}{P_2[i]+N_{22}}\right)\right]-\frac{n}{2}\log\left(2\pi e N_{22}\right)\nonumber\\
\stackrel{(c)}{\leq}& \frac{n}{2}\log\left[2\pi e\left( N_{21}+\frac{1}{n}\sum_{i=1}^n b'P_1[i]+\frac{\frac{1}{n}\sum_{i=1}^n P_2[i]N_{22}}{\frac{1}{n
}\sum_{i=1}^n P_2[i]+N_{22}}\right)\right]-\frac{n}{2}\log\left(2\pi e N_{22}\right)\nonumber\\
\stackrel{}{\leq}& \frac{n}{2}\log\left[2\pi e\left( N_{21}+b'P_1+\frac{P_2 N_{22}}{P_2+N_{22}}\right)\right]-\frac{n}{2}\log\left(2\pi e N_{22}\right)\nonumber\\
\stackrel{}{\leq}& \frac{n}{2}\log\left(
\frac{N_{21}}{N_{22}}+\frac{b'P_1}{N_{22}}+\frac{P_2
}{P_2+N_{22}}\right),
\end{IEEEeqnarray}
where (a) follows from the chain rule and the fact that removing
independent conditions does not decrease differential entropy, (b)
follows from the fact that Gaussian distribution maximizes the
conditional entropy for a given covariance matrix, and (c) follows
form Jenson's inequality.

For the last bracket, we again make use of the definition of $f_h$. In fact, since $aN_{22}\leq 1$, we have
\begin{IEEEeqnarray}{rl}
h(\tilde{y}_{22}^n)-\mu_1 h(\tilde{y}_{1}^n|x_1^n)&= h(x_2^n+z_{22}^n)-\mu_1 h(\sqrt{a}x_2^n+z_{1}^n)\nonumber\\
&\leq nf_h(P_2,N_{22},1,a,\mu_1).
\end{IEEEeqnarray}

Adding all inequalities, we obtain
\begin{IEEEeqnarray}{rl}\label{alaki1}
\mu R_1+R_2\leq &\frac{\mu_1}{2}\log\left[2\pi e(P_1+aP_2+1)\right]-\frac{\mu_2}{2}\log(2\pi e)+\frac{1}{2}\log\left( \frac{N_{21}}{N_{22}}+\frac{b'P_1}{N_{22}}+\frac{P_2 }{P_2+N_{22}}\right)\nonumber\\
&+\mu_2 f_h\left(P_1,1,N_{21},b',\frac{1}{\mu_2}\right)+f_h(P_2,N_{22},1,a,\mu_1),
\end{IEEEeqnarray}
where the fact that $\epsilon_n\rightarrow 0$ as $n\rightarrow
\infty$ is used  to eliminate $\epsilon_n$ form the right hand side
of the inequality. Now, by taking the minimum of the right hand side
of (\ref{alaki1}) over all $\mu_1$ and $\mu_2$, we obtain the
desired result. This completes the proof.
\end{proof}

\subsubsection{Class A2}
This class is the complement of Class A1 in the sense that we use it
to upper bound $\sigma_{\mathscr{C}}(1,\mu)$. A member of this class
is a channel in which User 1 is equipped with one transmit and two
receive antennas, whereas User 2 is equipped with one antenna at
both transmitter and receiver sides (see Figure \ref{classa2
admissible channels}). The channel model can be written as
\begin{equation}\label{class A2}
\begin{array}{rl}
\tilde{y}_{11}=&x_{1}+z_{11},\\
\tilde{y}_{12}=&x_{1}+\sqrt{a'}x_{2}+z_{12},\\
\tilde{y}_{2}=&x_{2}+\sqrt{b}x_{1}+z_{2},
\end{array}
\end{equation}
where $\tilde{y}_{11}$ and $\tilde{y}_{12}$ are the signals at the
first receiver, $\tilde{y}_{2}$ is the signal at the second
receiver, $z_{2}$ is additive Gaussian noise with unit variance,
$z_{11}$ and $z_{12}$ are additive Gaussian noise with variances
$N_{11}$ and $N_{12}$, respectively. Transmitter 1 and 2 are subject
to the power constraints $P_1$ and $P_2$, respectively.
\begin{figure}
\centering \includegraphics[scale=0.65]{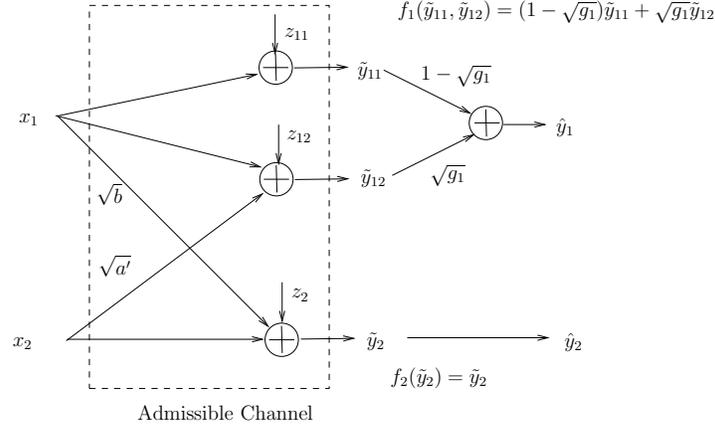}
\caption{Class A2 admissible channels.}\label{classa2 admissible channels}
\end{figure}

For this class, we consider two linear functions $f_1$ and $f_2$ as
follows (see Figure \ref{classa2 admissible channels}):
\begin{IEEEeqnarray}{rl}
f_1(\tilde{y}_{11}^n,\tilde{y}_{12}^n)&= (1-\sqrt{g_1})\tilde{y}_{11}^n+\sqrt{g_1}\tilde{y}_{12}^n,\\
f_2(\tilde{y}_{2}^n)&=\tilde{y}_{2}^n.
\end{IEEEeqnarray}
Similar to Class A1, when $g_1=0$, the admissible channels in Class
A2 become the one-sided Gaussian IC by letting $N_{12}\rightarrow
\infty$ and $N_{11}=1$. Therefore, we have
\begin{IEEEeqnarray}{rl}
\hat{y}_1^n=&x_1^n+\sqrt{a'g_1}x_2^n+(1-\sqrt{g_1})z_{11}^n+\sqrt{g_1}z_{12}^n,\\
\hat{y}_2^n=&\sqrt{b}x_1^n+x_2^n+z_{2}^n.
\end{IEEEeqnarray}

We conclude that the channel modeled by (\ref{class A2}) is
admissible if the corresponding parameters satisfy
\begin{equation}\label{class A2 conditions}
\begin{array}{rl}
a'g_1&=a,\\
(1-\sqrt{g_1})^2N_{11}+g_1N_{12}&=1.
\end{array}
\end{equation}
Similar to Class A1, we further add the following constraints to the
conditions of Class A2 channels:
\begin{equation}\label{class A2 conditions1}
\begin{array}{rl}
a'&\leq N_{12},\\
bN_{11}&\leq 1.
\end{array}
\end{equation}

In the following lemma, we obtain the required upper bound.

\begin{lemma}\label{class A2 lemma}
For the channels modeled by (\ref{class A2}) and satisfying
(\ref{class A2 conditions1}), we have
\begin{IEEEeqnarray}{rl}
\sigma_{\mathscr{C}'}(1,\mu)\leq&\min \frac{\mu_1}{2}\log\left[2\pi
e(bP_1+P_2+1)\right]-\frac{\mu_2}{2}\log(2\pi e)
+\frac{1}{2}\log\left( \frac{N_{12}}{N_{11}}+\frac{a'P_2}{N_{11}}+\frac{P_1 }{P_1+N_{11}}\right)\label{class A2 solution}\\
&~~~~~~~~~~~~+\mu_2 f_h\left(P_2,1,N_{12},a',\frac{1}{\mu_2}\right) +f_h(P_1,N_{11},1,b,\mu_1)\nonumber\\
&\text{subject to:}~\mu_1+\mu_2=\mu,~\mu_1,\mu_2\geq 0.\nonumber
\end{IEEEeqnarray}
\end{lemma}

\begin{proof}
The proof is similar to that of Lemma \ref{class A1 lemma} and we
omit it here.
\end{proof}

\subsubsection{Class B}
\begin{figure}
\centering \includegraphics[scale=0.65]{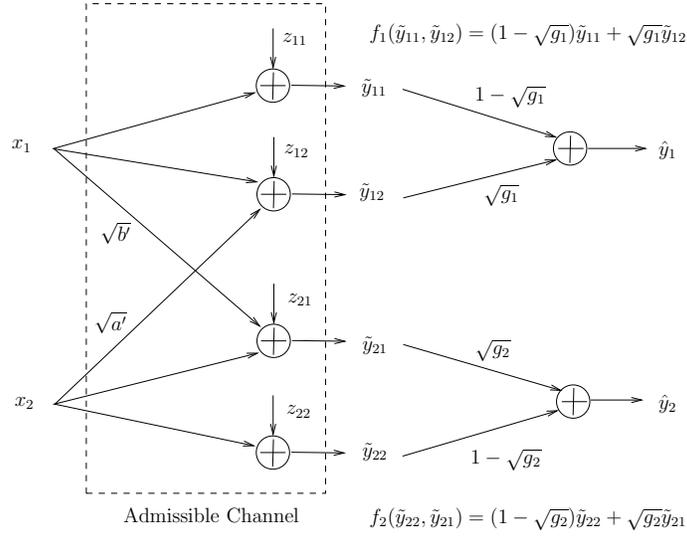}
\caption{Class B admissible channels.}\label{classb admissible channels}
\end{figure}

A member of this class is a channel with one transmit antenna and
two receive antennas for each user modeled by (see Figure
\ref{classb admissible channels})
\begin{equation}\label{class B}
\begin{array}{rl}
\tilde{y}_{11}=&x_{1}+z_{11},\\
\tilde{y}_{12}=&x_{1}+\sqrt{a'}x_{2}+z_{12},\\
\tilde{y}_{21}=&x_{2}+\sqrt{b'}x_{1}+z_{21},\\
\tilde{y}_{22}=&x_{2}+z_{22},
\end{array}
\end{equation}
where $\tilde{y}_{11}$ and $\tilde{y}_{12}$ are the signals at the
first receiver, $\tilde{y}_{21}$ and $\tilde{y}_{22}$ are the
signals at the second receiver, and $z_{ij}$ is additive Gaussian
noise with variance $N_{ij}$ for $i,j\in \{1,2\}$. Transmitter 1 and
2 are subject to the power constraints $P_1$ and $P_2$,
respectively. In fact, this channel is designed to upper bound both
$\sigma_{\mathscr{C}}(\mu,1)$ and $\sigma_{\mathscr{C}}(1,\mu)$.

Next, we investigate admissibility of this channel and the
conditions that must be imposed on the underlying parameters. Let us
consider two linear deterministic functions $f_1$ and $f_2$ with
parameters $0\leq g_1$ and $0\leq g_2$, respectively, as follows
(see Figure \ref{classb admissible channels})
\begin{IEEEeqnarray}{rl}
f_1(\tilde{y}_{11}^n,\tilde{y}_{12}^n)&=(1-\sqrt{g_1})\tilde{y}_{11}^n +\sqrt{g_1}\tilde{y}_{12}^n,\\
f_2(\tilde{y}_{22}^n,\tilde{y}_{21}^n)&= (1-\sqrt{g_2})\tilde{y}_{22}^n+\sqrt{g_2}\tilde{y}_{21}^n.
\end{IEEEeqnarray}
Therefore, we have
\begin{IEEEeqnarray}{rl}
\hat{y}_1^n=&x_1^n+\sqrt{a'g_1}x_2^n+(1-\sqrt{g_1})z_{11}^n+\sqrt{g_1}z_{12}^n,\\
\hat{y}_2^n=&\sqrt{b'g_2}x_1^n+x_2^n+(1-\sqrt{g_2})z_{22}^n+\sqrt{g_2}z_{21}^n.
\end{IEEEeqnarray}
To satisfy (\ref{condition3}) and (\ref{condition4}), it suffices to have
\begin{equation}\label{class B conditions}
\begin{array}{rl}
a'g_1&=a,\\
b'g_2&=b,\\
(1-\sqrt{g_1})^2N_{11}+g_1N_{12}&=1,\\
(1-\sqrt{g_2})^2N_{22}+g_2N_{21}&=1.
\end{array}
\end{equation}
Hence, a channel modeled by (\ref{class B}) is admissible if there
exist two nonnegative numbers $g_1$ and $g_2$ such that the
equalities in (\ref{class B conditions}) are satisfied. We further
add the following two constraints to the equality conditions in
(\ref{class B conditions}):
\begin{equation}\label{class B conditions1}
\begin{array}{rl}
b'N_{11}&\leq N_{21},\\
a'N_{22}&\leq N_{12}.
\end{array}
\end{equation}
Although adding more constraints reduces the number of the
admissible channels, it enables us to compute an outer bound on
$\sigma_{\mathscr{C}'}(\mu,1)$ and $\sigma_{\mathscr{C}'}(1,\mu)$.

\begin{lemma}\label{class B lemma}
For the channels modeled by (\ref{class B}) and satisfying (\ref{class B conditions1}), we have
\begin{IEEEeqnarray}{rl}
\sigma_{\mathscr{C}'}(\mu,1)\leq & \mu\gamma\left(\frac{P_1}{N_{11}}+\frac{P_1}{a'P_2+N_{12}}\right)
+\gamma\left(\frac{P_2}{N_{22}}+\frac{P_2}{b'P_1+N_{21}}\right)\nonumber\\
&+f_h(P_2,N_{22},N_{12},a',\mu)+\frac{\mu}{2}\log((2\pi e)(a'P_2+N_{12}))-\frac{1}{2}\log((2\pi e)(P_2+N_{22})),\label{class B solution1}\\
\sigma_{\mathscr{C}'}(1,\mu)\leq &
\gamma\left(\frac{P_1}{N_{11}}+\frac{P_1}{a'P_2+N_{12}}\right)+
\mu\gamma\left(\frac{P_2}{N_{22}}+\frac{P_2}{b'P_1+N_{21}}\right)
\nonumber\\
&+f_h(P_1,N_{11},N_{21},b',\mu)+\frac{\mu}{2}\log((2\pi e)(b'P_1+N_{21}))-\frac{1}{2}\log((2\pi e)(P_1+N_{11})).\label{class B solution2}
\end{IEEEeqnarray}

\end{lemma}
\begin{proof}
We only upper bound $\sigma_{\mathscr{C}'}(\mu,1)$ and an upper
bound on $\sigma_{\mathscr{C}'}(1,\mu)$ can be similarly obtained.
Let us assume $R_1$ and $R_2$ are achievable rates for User 1 and
User 2, respectively. Using Fano's inequality, we obtain
\begin{IEEEeqnarray}{rl}
n(\mu R_1+R_2)\stackrel{}{\leq}& \mu I(x_1^n;\tilde{y}_{11}^n,\tilde{y}_{12}^n)+ I(x_2^n;\tilde{y}_{22}^n,\tilde{y}_{21}^n)+n\epsilon_n\nonumber\\
\stackrel{}{=}& \mu I(x_1^n;\tilde{y}_{12}^n|\tilde{y}_{11}^n)+\mu I(x_1^n;\tilde{y}_{11}^n)\nonumber\\
&+I(x_2^n;\tilde{y}_{21}^n|\tilde{y}_{22}^n,)+I(x_2^n;\tilde{y}_{22}^n)+n\epsilon_n\nonumber\\
\stackrel{}{=}& \mu h(\tilde{y}_{12}^n|\tilde{y}_{11}^n)-\mu h(\tilde{y}_{12}^n|x_1^n, \tilde{y}_{11}^n)+\mu h(\tilde{y}_{11}^n)-\mu h(\tilde{y}_{11}^n|x_1^n)\nonumber\\
&+h(\tilde{y}_{21}^n|\tilde{y}_{22}^n)-h(\tilde{y}_{21}^n|x_2^n,\tilde{y}_{22}^n)+h(\tilde{y}_{22}^n)-h(\tilde{y}_{22}^n|x_2^n)+n\epsilon_n\nonumber\\
\stackrel{}{=}& \big[\mu h(\tilde{y}_{12}^n|\tilde{y}_{11}^n)-\mu h(\tilde{y}_{11}^n|x_1^n)\big] +\big[h(\tilde{y}_{21}^n|\tilde{y}_{22}^n)-h(\tilde{y}_{22}^n|x_2^n)\big]\nonumber\\
&+\big[\mu h(\tilde{y}_{11}^n)-h(\tilde{y}_{21}^n|x_2^n,
\tilde{y}_{22}^n)\big]+ \big[h(\tilde{y}_{22}^n)-\mu
h(\tilde{y}_{12}^n|x_1^n,\tilde{y}_{11}^n)\big]
+n\epsilon_n.\label{class B fano}
\end{IEEEeqnarray}

Next, we upper bound the terms within each bracket in (\ref{class B
fano}) separately. For the first bracket, we have
\begin{IEEEeqnarray}{rl}
\mu h(\tilde{y}_{12}^n|\tilde{y}_{11}^n)-\mu h(\tilde{y}_{11}^n|x_1^n)\stackrel{(a)}{\leq}& \mu\sum_{i=1}^n h(\tilde{y}_{12}[i]|\tilde{y}_{11}[i])-\frac{\mu n}{2}\log\left(2\pi e N_{11}\right)\nonumber\\
\stackrel{(b)}{\leq}& \mu\sum_{i=1}^n\frac{1}{2}\log\left[2\pi e\left( N_{12}+a'P_2[i]+\frac{P_1[i]N_{11}}{P_1[i]+N_{11}}\right)\right]-\frac{\mu n}{2}\log\left(2\pi e N_{11}\right)\nonumber\\
\stackrel{(c)}{\leq}& \frac{\mu n}{2}\log\left[2\pi e\left( N_{12}+\frac{1}{n}\sum_{i=1}^n a'P_2[i]+\frac{\frac{1}{n}\sum_{i=1}^n P_1[i]N_{11}}{\frac{1}{n
}\sum_{i=1}^n P_1[i]+N_{11}}\right)\right]-\frac{\mu n}{2}\log\left(2\pi e N_{11}\right)\nonumber\\
\stackrel{}{\leq}& \frac{\mu n}{2}\log\left[2\pi e\left( N_{12}+a'P_2+\frac{P_1 N_{11}}{P_1+N_{11}}\right)\right]-\frac{\mu n}{2}\log\left(2\pi e N_{11}\right)\nonumber\\
\stackrel{}{=}& \frac{\mu n}{2}\log\left(
\frac{N_{12}}{N_{11}}+\frac{a'P_2}{N_{11}}+\frac{P_1}{P_1+N_{11}}\right),
\end{IEEEeqnarray}
where (a) follows from the chain rule and the fact that removing
independent conditions increases differential entropy, (b) follows
from the fact that Gaussian distribution optimizes conditional
entropy for a given covariance matrix, and (c) follows form Jenson's
inequality.

Similarly, the terms within the second bracket can be upper bounded
as
\begin{equation}
h(\tilde{y}_{21}^n|\tilde{y}_{22}^n)-h(\tilde{y}_{22}^n|x_2^n)\leq
\frac{n}{2}\log\left(
\frac{N_{21}}{N_{22}}+\frac{b'P_1}{N_{22}}+\frac{P_2}{P_2+N_{22}}\right).
\end{equation}

Using Lemma \ref{lem Extremal-1} and the fact that $N_{11}\leq
N_{21}/b'$, the terms within the third bracket can be upper bounded
as
\begin{IEEEeqnarray}{rl}
\mu h(\tilde{y}_{11}^n)-h(\tilde{y}_{21}^n|x_2^n,\tilde{y}_{22}^n)&= \mu \left(h(x_1^n+z_{11}^n)-\frac{1}{\mu}h(\sqrt{b'}x_1^n+z_{21}^n)\right)\nonumber\\
&\leq \mu nf_h\left(P_1,N_{11},N_{21},b',\frac{1}{\mu}\right).
\end{IEEEeqnarray}
Since $1\leq\mu$, from (\ref{extremal-function}) we obtain
\begin{equation}
\mu
h(\tilde{y}_{11}^n)-h(\tilde{y}_{21}^n|x_2^n,\tilde{y}_{22}^n)\leq
\frac{\mu n}{2}\log((2\pi e)(P_1+N_{11}))-\frac{n}{2}\log((2\pi
e)(b'P_1+N_{21})).
\end{equation}

For the last bracket, again we use Lemma \ref{lem Extremal-1} to
obtain
\begin{IEEEeqnarray}{rl}
h(\tilde{y}_{22}^n)-\mu h(\tilde{y}_{12}^n|x_1^n,\tilde{y}_{11}^n)&= h(x_2^n+z_{22}^n)-\mu h(\sqrt{a'}x_2^n+z_{12}^n)\nonumber\\
&\leq nf_h(P_2,N_{22},N_{12},a',\mu).
\end{IEEEeqnarray}

Adding all inequalities, we have
\begin{IEEEeqnarray}{rl}
\mu R_1+R_2\leq &\frac{\mu}{2}\log\left( \frac{N_{12}}{N_{11}}+\frac{a'P_2}{N_{11}}+\frac{P_1}{P_1+N_{11}}\right)+ \frac{1}{2}\log\left( \frac{N_{21}}{N_{22}}+\frac{b'P_1}{N_{22}}+\frac{P_2}{P_2+N_{22}}\right)
\nonumber \\
&+\frac{\mu}{2}\log((2\pi e)(P_1+N_{11}))-\frac{1}{2}\log((2\pi e)(b'P_1+N_{21}))+f_h(P_2,N_{22},N_{12},a',\mu),
\end{IEEEeqnarray}
where the fact that $\epsilon_n\rightarrow 0$ as $n\rightarrow
\infty$ is used to eliminate $\epsilon_n$ from the right hand side
of the inequality. By rearranging the terms, we obtain
\begin{IEEEeqnarray}{rl}
\mu R_1+R_2\leq & \mu\gamma\left(\frac{P_1}{N_{11}}+\frac{P_1}{a'P_2+N_{12}}\right)
+\gamma\left(\frac{P_2}{N_{22}}+\frac{P_2}{b'P_1+N_{21}}\right)\nonumber\\
&+f_h(P_2,N_{22},N_{12},a',\mu)+\frac{\mu}{2}\log((2\pi e)(a'P_2+N_{12}))-\frac{1}{2}\log((2\pi e)(P_2+N_{22})).\nonumber
\end{IEEEeqnarray}
This completes the proof.
\end{proof}

A unique feature of the channels within Class B is that for
$1\leq\mu\leq \frac{P_2+N_{12}/a'}{P_2+N_{22}}$ and $1\leq\mu\leq
\frac{P_1+N_{21}/b'}{P_1+N_{11}}$, the upper bounds in (\ref{class B
solution1}) and (\ref{class B solution2}) become, respectively,
\begin{IEEEeqnarray}{rl}\label{alaki2}
\mu R_1+R_2\leq & \mu\gamma\left(\frac{P_1}{N_{11}}+\frac{P_1}{a'P_2+N_{12}}\right)
+\gamma\left(\frac{P_2}{N_{22}}+\frac{P_2}{b'P_1+N_{21}}\right)
\end{IEEEeqnarray}
and
\begin{IEEEeqnarray}{rl}
R_1+\mu R_2\leq & \gamma\left(\frac{P_1}{N_{11}}+\frac{P_1}{a'P_2+N_{12}}\right)+ \mu\gamma\left(\frac{P_2}{N_{22}}+\frac{P_2}{b'P_1+N_{21}}\right).
\end{IEEEeqnarray}
On the other hand, if the receivers treat the interference as noise,
it can be shown that
\begin{equation}
R_1= \gamma\left(\frac{P_1}{N_{11}}+\frac{P_1}{a'P_2+N_{12}}\right)
\end{equation}
and
\begin{equation}
R_2= \gamma\left(\frac{P_2}{N_{22}}+\frac{P_2}{b'P_1+N_{21}}\right)
\end{equation}
are achievable. Comparing upper bounds and achievable rates, we
conclude that the upper bounds are indeed tight. In fact, this
property is first observed by Etkin {\em et al.} in \cite{Etkin:IC}.
We summarize this result in the following theorem:

\begin{theorem}\label{class B theorem sum capacity}
The sum capacity in Class B is attained when transmitters use
Gaussian codebooks and receivers treat the interference as noise. In
this case, the sum capacity is
\begin{IEEEeqnarray}{rl}
\mathscr{C}'_{\text{sum}}= & \gamma\left(\frac{P_1}{N_{11}}+\frac{P_1}{a'P_2+N_{12}}\right)
+\gamma\left(\frac{P_2}{N_{22}}+\frac{P_2}{b'P_1+N_{21}}\right).
\end{IEEEeqnarray}

\end{theorem}

\begin{proof}
By substituting $\mu=1$ in (\ref{alaki2}), we obtain the desired
result.
\end{proof}

\subsubsection{Class C}
Class C is designed to upper bound $\sigma_{\mathscr{C}}(\mu,1)$ for
the mixed Gaussian IC where $1\leq b$. Class C is similar to Class
A1 (see Figure \ref{classa1 admissible channels}), however we impose
different constraints on the parameters of the channels within Class
C. These constraints assist us in providing upper bounds by using
the fact that at one of the receivers both signals are decodable.

For channels in Class C, we use the same model that is given in
(\ref{class A1}). Therefore, similar to channels in Class A1, this
channel is admissible if the corresponding parameters satisfy
\begin{equation}\label{class C conditions}
\begin{array}{rl}
b'g_2&=b,\\
(1-\sqrt{g_2})^2N_{22}+g_2N_{21}&=1.
\end{array}
\end{equation}
Next, we change the constraints in (\ref{class A1 conditions1}) as
\begin{equation}\label{class C conditions1}
\begin{array}{rl}
b'&\geq N_{21},\\
aN_{22}&\leq 1.
\end{array}
\end{equation}
Through this change of constraints, the second receiver after
decoding its own signal will have a less noisy version of the first
user's signal, and consequently, it is able to decode the signal of
the first user as well as its own signal. Relying on this
observation, we have the following lemma.

\begin{lemma}\label{class C lemma}
For a channel in Class C, we have
\begin{IEEEeqnarray}{rl}
\sigma_{\mathscr{C}'}(\mu,1)\leq & \frac{\mu-1}{2}\log\left(2\pi e(P_1+aP_2+1)\right)+\frac{1}{2}\log\left(2\pi e \left(\frac{P_2N_{22}}{P_2+N_{22}}+b'P_1+N_{21}\right)\right)\nonumber\\
&~~~~-\frac{1}{2}\log(2\pi e N_{21})-\frac{1}{2}\log(2\pi e N_{22})+
f_h(P_2,N_{22},1,a,\mu-1).
\end{IEEEeqnarray}

\end{lemma}

\begin{proof}
Since the second user is able to decode both users' messages, we have
\begin{IEEEeqnarray}{rl}
R_1&\leq \frac{1}{n}I(x_1^n;\tilde{y}_1^n),\\
R_1&\leq \frac{1}{n}I(x_1^n;\tilde{y}_{21}^n,\tilde{y}_{22}^n|x_2^n),\label{class C alaki}\\
R_2&\leq \frac{1}{n}I(x_2^n;\tilde{y}_{21}^n,\tilde{y}_{22}^n|x_1^n),\\
R_1+R_2&\leq
\frac{1}{n}I(x_1^n,x_2^n;\tilde{y}_{21}^n,\tilde{y}_{22}^n).
\end{IEEEeqnarray}
From $aN_{22}\leq 1$, we have $I(x_1^n;\tilde{y}_1^n)\leq
I(x_1^n;\tilde{y}_{21}^n|x_2^n)
=I(x_1^n;\tilde{y}_{21}^n,\tilde{y}_{22}^n|x_2^n)$. Hence,
(\ref{class C alaki}) is redundant. It can be shown that
\begin{equation}
\mu R_1+R_2\leq \frac{\mu-1}{n}I(x_1^n;\tilde{y}_1^n)+
\frac{1}{n}I(x_1^n,x_2^n;\tilde{y}_{21}^n,\tilde{y}_{22}^n).
\end{equation}
Hence, we have
\begin{IEEEeqnarray}{rl}
\nonumber\mu R_1+R_2&\leq \frac{\mu-1}{n}h(\tilde{y}_1^n)-\frac{\mu-1}{n}h(\tilde{y}_1^n|x_1^n)+ \frac{1}{n}h(\tilde{y}_{21}^n,\tilde{y}_{22}^n) -\frac{1}{n}h(\tilde{y}_{21}^n,\tilde{y}_{22}^n|x_1^n,x_2^n)\\
\nonumber&= \frac{\mu-1}{n}h(\tilde{y}_1^n)+ \frac{1}{n}h(\tilde{y}_{21}^n|\tilde{y}_{22}^n) -\frac{1}{n}h(\tilde{y}_{21}^n,\tilde{y}_{22}^n|x_1^n,x_2^n)\\
\label{class C alaki1}&~~+\left[\frac{1}{n}h(\tilde{y}_{22}^n)- \frac{\mu-1}{n}h(\tilde{y}_1^n|x_1^n)\right]
\end{IEEEeqnarray}
Next, we bound the different terms in (\ref{class C alaki1}). For
the first term, we have
\begin{equation}
\frac{\mu-1}{n}h(\tilde{y}_1^n)\leq \frac{\mu-1}{2}\log\left(2\pi e(P_1+aP_2+1)\right).
\end{equation}
The second term can be bounded as
\begin{equation}
\frac{1}{n}h(\tilde{y}_{21}^n|\tilde{y}_{22}^n)\leq \frac{1}{2}\log\left(2\pi e \left(\frac{P_2N_{22}}{P_2+N_{22}}+b'P_1+N_{21}\right)\right).
\end{equation}
The third term can be bounded as
\begin{equation}
\frac{1}{n}h(\tilde{y}_{21}^n,\tilde{y}_{22}^n|x_1^n,x_2^n)=\frac{1}{2}\log(2\pi
e N_{21})+\frac{1}{2}\log(2\pi e N_{22}).
\end{equation}
The last terms can be bounded as
\begin{IEEEeqnarray}{rl}
\frac{1}{n}h(\tilde{y}_{22}^n)- \frac{\mu-1}{n}h(\tilde{y}_1^n|x_1^n)&=\frac{1}{n}h(x_2^n+z_{22}^n)- \frac{\mu-1}{n}h(\sqrt{a}x_2^n+z_1)\\
&\leq f_h(P_2,N_{22},1,a,\mu-1).
\end{IEEEeqnarray}
Adding all inequalities, we obtain the desired result.
\end{proof}

\section{Weak Gaussian Interference Channel}
In this section, we focus on the weak Gaussian IC. We first obtain
the sum capacity of this channel for a certain range of parameters.
Then, we obtain an outer bound on the capacity region which is
tighter than the previously known outer bounds. Finally, we show
that time-sharing and concavification result in the same achievable
region for Gaussian codebooks.

\subsection{Sum Capacity}
In this subsection, we use the Class B channels to obtain the sum
capacity of the weak IC for a certain range of parameters. To this
end, let us consider the following minimization problem:
\begin{IEEEeqnarray}{rl}
W=&\min \gamma\left(\frac{P_1}{N_{11}}+\frac{P_1}{a'P_2+N_{12}}\right)
+\gamma\left(\frac{P_2}{N_{22}}+\frac{P_2}{b'P_1+N_{21}}\right)\label{sum capacity-objective}\\
&\text{subject to:}\nonumber\\
&~~~~~~~a'g_1=a\nonumber\\
&~~~~~~~b'g_2=b\nonumber\\
&~~~~~~~b'N_{11}\leq N_{21}\nonumber\\
&~~~~~~~a'N_{22}\leq N_{12}\nonumber\\
&~~~~~~~(1-\sqrt{g_1})^2N_{11}+g_1N_{12}=1\nonumber\\
&~~~~~~~(1-\sqrt{g_2})^2N_{22}+g_2N_{21}=1\nonumber\\
&~~~~~~~0\leq [a',b',g_1,g_2,N_{11},N_{12},N_{22},N_{21}].\nonumber
\end{IEEEeqnarray}
The objective function in (\ref{sum capacity-objective}) is the sum
capacity of Class B channels obtained in Theorem \ref{class B
theorem sum capacity}. The constraints are the combination of
(\ref{class B conditions}) and (\ref{class B conditions1}) where
applied to confirm the admissibility of the channel and to validate
the sum capacity result. Since every channel in the class is
admissible, we have $\mathscr{C}_{sum}\leq W$. Substituting
$S_1=g_1N_{12}$ and $S_2=g_2N_{21}$, we have
\begin{IEEEeqnarray}{rl}
W=&\min \gamma\left(\frac{(1-\sqrt{g_1})^2P_1}{1-S_1}+\frac{g_1P_1}{aP_2+S_{1}}\right)
+\gamma\left(\frac{(1-\sqrt{g_2})^2P_2}{1-S_{2}}+\frac{g_2P_2}{bP_1+S_{2}}\right)\label{sum
capacity-objective1}\\
&\text{subject to:}\nonumber\\
&~~~~~~~\frac{b(1-S_1)}{(1-\sqrt{g_1})^2}\leq S_2< 1\nonumber\\
&~~~~~~~\frac{a(1-S_2)}{(1-\sqrt{g_2})^2}\leq S_1< 1\nonumber\\
&~~~~~~~0< [g_1,g_2].\nonumber
\end{IEEEeqnarray}

By first minimizing with respect to $g_1$ and $g_2$, the
optimization problem (\ref{sum capacity-objective1}) can be
decomposed as
\begin{IEEEeqnarray}{rl}
W=&\min W_1+W_2\\
&\text{subject to:}~0<S_1< 1,~0<S_2< 1.\nonumber
\end{IEEEeqnarray}
where $W_1$ is defined as
\begin{IEEEeqnarray}{rl}
W_1=&\min_{g_1} \gamma\left(\frac{(1-\sqrt{g_1})^2P_1}{1-S_1}+\frac{g_1P_1}{aP_2+S_{1}}\right)\label{optim1}\\
&\text{subject to:}~\frac{b(1-S_1)}{S_2}\leq (1-\sqrt{g_1})^2,~0<
g_1.\nonumber
\end{IEEEeqnarray}
Similarly, $W_2$ is defined as
\begin{IEEEeqnarray}{rl}
W_2=&\min_{g_2} \gamma\left(\frac{(1-\sqrt{g_2})^2P_2}{1-S_2}+\frac{g_2P_2}{bP_1+S_{2}}\right)\label{optim2}\\
&\text{subject to:}~\frac{a(1-S_2)}{S_1}\leq (1-\sqrt{g_2})^2,~0<
g_2.\nonumber
\end{IEEEeqnarray}
The optimization problems (\ref{optim1}) and (\ref{optim2}) are easy
to solve. In fact, we have
\begin{equation}\label{opt1}
W_1=\left\{
\begin{array}{ll}
\gamma\left(\frac{P_1}{1+aP_2}\right)& ~\text{if} ~ \sqrt{b}(1+aP_2)\leq \sqrt{S_2(1-S_1)}\\
\gamma\left(\frac{bP_1}{S_2}+\frac{(1-\sqrt{b(1-S_1)/S_2})^2P_1}{aP_2+S_1}\right)& ~\text{Otherwise}
\end{array}
\right.
\end{equation}
\begin{equation}\label{opt2}
W_2=\left\{
\begin{array}{ll}
\gamma\left(\frac{P_2}{1+bP_1}\right)& ~\text{if} ~ \sqrt{a}(1+bP_1)\leq \sqrt{S_1(1-S_2)}\\
\gamma\left(\frac{aP_2}{S_1}+\frac{(1-\sqrt{a(1-S_2)/S_1})^2P_2}{bP_1+S_2}\right)& ~\text{Otherwise}
\end{array}
\right.
\end{equation}

From (\ref{opt1}) and (\ref{opt2}), we observe that for $S_1$ and
$S_2$ satisfying $\sqrt{b}(1+aP_2)\leq \sqrt{S_2(1-S_1)}$ and
$\sqrt{a}(1+bP_1)\leq \sqrt{S_1(1-S_2)}$, the objective function
becomes independent of $S_1$ and $S_2$. In this case, we have
\begin{equation}\label{s1s2}
W=\gamma\left(\frac{P_1}{1+aP_2}\right)+\gamma\left(\frac{P_2}{1+bP_1}\right),
\end{equation}
which is achievable by treating interference as noise. In the
following theorem, we prove that it is possible to find a certain
range of parameters such that there exist $S_1$ and $S_2$ yielding
(\ref{s1s2}).

\begin{theorem}
The sum capacity of the two-user Gaussian IC is
\begin{equation}
\mathscr{C}_{sum}=\gamma\left(\frac{P_1}{1+aP_2}\right)
+\gamma\left(\frac{P_2}{1+bP_1}\right),
\end{equation}
for the range of parameters satisfying
\begin{equation}\label{constraint}
\sqrt{b}P_1+\sqrt{a}P_2\leq \frac{1-\sqrt{a}-\sqrt{b}}{\sqrt{ab}}.
\end{equation}
\end{theorem}

\begin{proof}
Let us fix $a$ and $b$, and define $D$ as
\begin{equation}\label{D region}
D=\left\{(P_1,P_2)|P_1\leq
\frac{\sqrt{S_1(1-S_2)}}{b\sqrt{a}}-\frac{1}{b},
P_2\leq\frac{\sqrt{S_2(1-S_1)}}{a\sqrt{b}}-\frac{1}{a},0<S_1<1,0<S_2<1\right\}.
\end{equation}
In fact, if $D$ is feasible then there exist $0<S_1<1$ and $0<S_2<1$
satisfying $\sqrt{b}(1+aP_2)\leq \sqrt{S_2(1-S_1)}$ and
$\sqrt{a}(1+bP_1)\leq \sqrt{S_1(1-S_2)}$. Therefore, the sum
capacity of the channel for all feasible points is attained due to
(\ref{s1s2}).

We claim that $D=D'$, where $D'$ is defined as
\begin{equation}\label{D' region}
D'=\left\{(P_1,P_2)|\sqrt{b}P_1+\sqrt{a}P_2\leq \frac{1-\sqrt{a}-\sqrt{b}}{\sqrt{ab}}\right\}.
\end{equation}
To show $D'\subseteq D$, we set $S_1=1-S_2$ in (\ref{D region}) to get
\begin{equation}
\left\{(P_1,P_2)|P_1\leq \frac{S_1}{b\sqrt{a}}-\frac{1}{b}, P_2\leq\frac{1-S_1}{a\sqrt{b}}-\frac{1}{a},0<S_1<1\right\}\subseteq D.
\end{equation}
It is easy to show that the left hand side of the above equation is
another representation of the region $D'$. Hence, we have
$D'\subseteq D$.

To show $D\subseteq D'$, it suffices to prove that for any
$(P_1,P_2)\in D$, $\sqrt{b}P_1+\sqrt{a}P_2\leq
\frac{1-\sqrt{a}-\sqrt{b}}{\sqrt{ab}}$ holds. To this end, we
introduce the following maximization problem:
\begin{equation}
J=\max_{(P_1,P_2)\in D}\sqrt{b}P_1+\sqrt{a}P_2,
\end{equation}
which can be written as
\begin{equation}
J=\max_{(S_1,S_2)\in (0,1)^2}\frac{\sqrt{S_1(1-S_2)}+\sqrt{S_2(1-S_1)}}{\sqrt{ab}} -\frac{1}{\sqrt{a}}-\frac{1}{\sqrt{b}}.
\end{equation}
It is easy to show that the solution to the above optimization problem is \begin{equation}
J=\frac{1}{\sqrt{ab}} -\frac{1}{\sqrt{a}}-\frac{1}{\sqrt{b}}.
\end{equation}
Hence, we deduce that $D\subseteq D'$. This completes the proof.
\end{proof}

\begin{remark}
The above sum capacity result for the weak Gaussian IC (see also
\cite{abolfazl2}) has been established independently in \cite{shang}
and \cite{Annapureddy}.
\end{remark}

\begin{figure}
\centering \includegraphics[scale=.75]{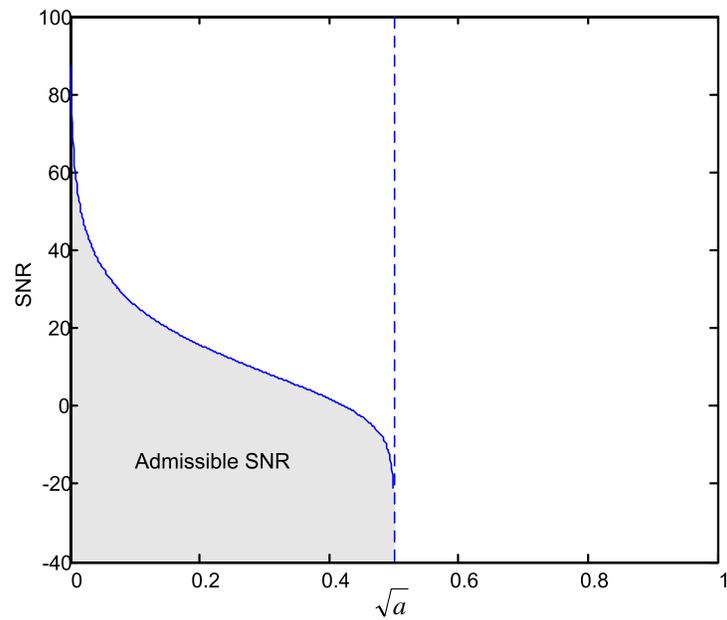}
\caption{The shaded area is the region where treating interference
as noise is optimal for obtaining the sum capacity of the symmetric
Gaussian IC.}\label{symmetric constraint}
\end{figure}

\begin{figure}
\centering \includegraphics[scale=0.75]{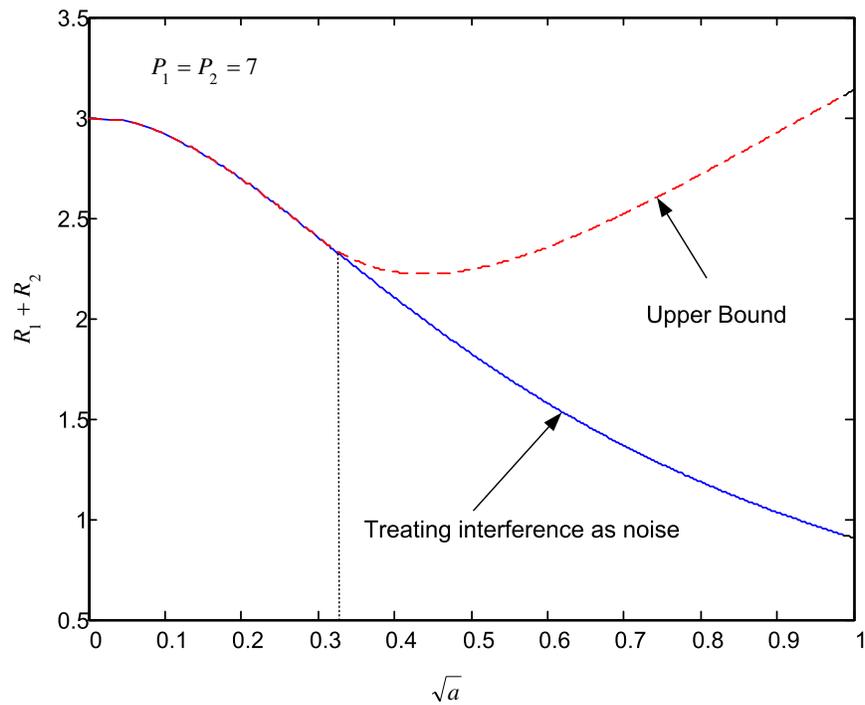}
\caption{The upper bound obtained by solving (\ref{sum
capacity-objective}). The lower bound is obtained by treating the
interference as noise.}\label{symmetric-noise}
\end{figure}

As an example, let us consider the symmetric Gaussian IC. In this
case, the constraint in (\ref{constraint}) becomes
\begin{equation}
P\leq \frac{1-2\sqrt{a}}{2a\sqrt{a}}.
\end{equation}
In Figure \ref{symmetric constraint}, the admissible region for $P$,
where treating interference as noise is optimal, versus $\sqrt{a}$
is plotted.
For a fixed $P$ and all $0\leq a\leq 1$, the upper bound in
(\ref{sum capacity-objective}) and the lower bound when receivers
treat the interference as noise are plotted in Figure
\ref{symmetric-noise}. We observe that up to a certain value of $a$,
the upper bound coincides with the lower bound.

\subsection{New Outer Bound}
For the weak Gaussian IC, there are two outer bounds that are
tighter than the other known bounds. The first one, due to Kramer
\cite{BOUNDS:Kramer}, is obtained by relying on the fact that the
capacity region of the Gaussian IC is inside the capacity regions of
the two underlying one-sided Gaussian ICs. Even though the capacity
region of the one-sided Gaussian IC is unknown, there exists an
outer bound for this channel that can be used instead. Kramers'
outer bound is the intersection of two regions $E_1$ and $E_2$.
$E_1$ is the collection of all rate pairs $(R_1,R_2)$ satisfying
\begin{IEEEeqnarray}{rl}
R_1&\leq \gamma\left(\frac{(1-\beta)P'}{\beta P'+1/a}\right),\\
R_2&\leq \gamma(\beta P'),
\end{IEEEeqnarray}
for all $\beta\in[0,\beta_{\text{max}}]$, where $P'=P_1/a+P_2$ and
$\beta_{\text{max}}=\frac{P_2}{P'(1+P_1)}$. Similarly, $E_2$ is the
collection of all rate pairs $(R_1,R_2)$ satisfying
\begin{IEEEeqnarray}{rl}
R_1&\leq \gamma(\alpha P''),\\
R_2&\leq \gamma\left(\frac{(1-\alpha)P''}{\alpha P''+1/b}\right),
\end{IEEEeqnarray}
for all $\alpha\in[0,\alpha_{\text{max}}]$, where $P''=P_1+P_2/b$
and $\alpha_{\text{max}}=\frac{P_1}{P''(1+P_2)}$.

The second outer bound, due to Etkin {\em et al.} \cite{Etkin:IC},
is obtained by using Genie aided technique to upper bound different
linear combinations of rates that appear in the HK achievable
region. Their outer bound is the union of all rate pairs $(R_1,R_2)$
satisfying
\begin{IEEEeqnarray}{rl}
R_1&\leq \gamma(P_1),\\
R_2&\leq \gamma(P_2),\\
R_1+R_2&\leq \gamma(P_1)+\gamma\left(\frac{P_2}{1+bP_1}\right),\\
R_1+R_2&\leq \gamma(P_2)+\gamma\left(\frac{P_1}{1+aP_2}\right),\\
R_1+R_2&\leq \gamma\left(aP_2+\frac{P_1}{1+bP_1}\right) +\gamma\left(bP_1+\frac{P_2}{1+aP_2}\right),\\
2R_1+R_2&\leq \gamma(P_1+aP_2)+\gamma\left(bP_1+\frac{P_2}{1+aP_2}\right)+ 0.5\log\left(\frac{1+P_1}{1+bP_1}\right),\\
R_1+2R_2&\leq \gamma(bP_1+P_2)+\gamma\left(aP_2+\frac{P_1}{1+bP_1}\right)+ 0.5\log\left(\frac{1+P_2}{1+aP_2}\right).
\end{IEEEeqnarray}

In the outer bound proposed here, we derive an upper bound on all
linear combinations of the rates. Recall that to obtain the boundary
points of the capacity region $\mathscr{C}$, it suffices to
calculate $\sigma_{\mathscr{C}}(\mu,1)$ and
$\sigma_{\mathscr{C}}(1,\mu)$ for all $1\leq\mu$. To this end, we
make use of channels in A1 and B classes and channels in A2 and B
classes to obtain upper bounds on $\sigma_{\mathscr{C}}(\mu,1)$ and
$\sigma_{\mathscr{C}}(1,\mu)$, respectively.

In order to obtain an upper bound on $\sigma_{\mathscr{C}}(\mu,1)$,
we introduce two optimization problems as follows. The first
optimization problem is written as
\begin{IEEEeqnarray}{rl}
W_1(\mu)=&\min \frac{\mu_1}{2}\log\left[2\pi e(P_1+aP_2+1)\right]-\frac{\mu_2}{2}\log(2\pi e)+\frac{1}{2}\log\left( \frac{N_{21}}{N_{22}}+\frac{b'P_1}{N_{22}}+\frac{P_2 }{P_2+N_{22}}\right)\label{Outer bound obj1}\\
&~~~~~~~~~~~~+\mu_2 f_h\left(P_1,1,N_{21},b',\frac{1}{\mu_2}\right) +f_h(P_2,N_{22},1,a,\mu_1)\nonumber\\
&\text{subject to:}\nonumber\\
&~~~~~~~\mu_1+\mu_2=\mu\nonumber\\
&~~~~~~~b'g_2=b\nonumber\\
&~~~~~~~b'\leq N_{21}\nonumber\\
&~~~~~~~aN_{22}\leq 1\nonumber\\
&~~~~~~~(1-\sqrt{g_2})^2N_{22}+g_2N_{21}=1\nonumber\\
&~~~~~~~0\leq [\mu_1,\mu_2,b',g_2,N_{22},N_{21}].\nonumber
\end{IEEEeqnarray}
In fact, the objective of the above minimization problem is an upper
bound on the support function of a channel within Class A1 which is
obtained in Lemma \ref{class A1 lemma}. The constraints are the
combination of (\ref{class A1 conditions}) and (\ref{class A1
conditions1}) which are applied to guarantee the admissibility of
the channel and to validate the upper bound obtained in Lemma
\ref{class A1 lemma}. Hence, $\sigma_{\mathscr{C}}(\mu,1)\leq
W_1(\mu)$. By using a new variable $S=(1-\sqrt{g_2})^2N_{22}$, we
obtain

\begin{IEEEeqnarray}{rl}
W_1(\mu)=&\min \frac{\mu_1}{2}\log\left[2\pi e(P_1+aP_2+1)\right]+\frac{1}{2}\log\left[(1-\sqrt{g_2})^2( \frac{1-S+bP_1}{g_2S}+\frac{P_2 }{(1-\sqrt{g_2})^2P_2+S})\right]\\
&~~~~~~~~~~~~+\mu_2 f_h\left(P_1,1,\frac{1-S}{g_2},\frac{b}{g_2},\frac{1}{\mu_2}\right) +f_h(P_2,\frac{S}{(1-\sqrt{g_2})^2},1,a,\mu_1)-\frac{\mu_2}{2}\log(2\pi e)\nonumber\\
&\text{subject to:}\nonumber\\
&~~~~~~~\mu_1+\mu_2=\mu\nonumber\\
&~~~~~~~S\leq 1-b\nonumber\\
&~~~~~~~S\leq \frac{(1-\sqrt{g_2})^2}{a}\nonumber\\
&~~~~~~~0\leq [\mu_1,\mu_2,S,g_2].\nonumber
\end{IEEEeqnarray}

The second optimization problem is written as
\begin{IEEEeqnarray}{rl}
W_2(\mu)=&\min \mu\gamma\left(\frac{P_1}{N_{11}}+\frac{P_1}{a'P_2+N_{12}}\right)
+\gamma\left(\frac{P_2}{N_{22}}+\frac{P_2}{b'P_1+N_{21}}\right) +f_h(P_2,N_{22},N_{12},a',\mu)\label{Outer bound obj2}\\
&~~~~~~~~~~~~~+\frac{\mu}{2}\log((2\pi e)(a'P_2+N_{12}))-\frac{1}{2}\log((2\pi e)(P_2+N_{22}))\nonumber\\
&\text{subject to:}\nonumber\\
&~~~~~~~a'g_1=a\nonumber\\
&~~~~~~~b'g_2=b\nonumber\\
&~~~~~~~b'N_{11}\leq N_{21}\nonumber\\
&~~~~~~~a'N_{22}\leq N_{12}\nonumber\\
&~~~~~~~(1-\sqrt{g_1})^2N_{11}+g_1N_{12}=1\nonumber\\
&~~~~~~~(1-\sqrt{g_2})^2N_{22}+g_2N_{21}=1\nonumber\\
&~~~~~~~0\leq [a',b',g_1,g_2,N_{11},N_{12},N_{22},N_{21}].\nonumber
\end{IEEEeqnarray}
For this problem, Class B channels are used. In fact, the objective
is the upper bound on the support function of channels within the
class obtained in Lemma \ref{class B lemma} and the constraints are
defined to obtain the closed form formula for the upper bound and to
confirm that the channels are admissible. Hence, we deduce
$\sigma_{\mathscr{C}}(\mu,1)\leq W_2(\mu)$. By using new variables
$S_1=g_1N_{12}$ and $S_2=g_2N_{21}$ , we obtain

\begin{IEEEeqnarray}{rl}
W_2(\mu)=&\min \mu \gamma\left(\frac{(1-\sqrt{g_1})^2P_1}{1-S_1}
+\frac{g_1P_1}{aP_2+S_{1}}\right)
+\gamma\left(\frac{(1-\sqrt{g_2})^2P_2}{1-S_{2}}+\frac{g_2P_2}{bP_1+S_{2}}\right)\\
&~~~~~~~~~~~~~+f_h\left(P_2,\frac{1-S_2}{(1-\sqrt{g_2})^2},\frac{S_1}{g_1},\frac{a}{g_1},\mu\right) +\frac{\mu}{2}\log\left((2\pi e)(\frac{aP_2+S_1}{g_1})\right)-\frac{1}{2}\log\left((2\pi e)(P_2+\frac{1-S_2}{(1-\sqrt{g_2})^2})\right)\nonumber\\
&\text{subject to:}\nonumber\\
&~~~~~~~\frac{b(1-S_1)}{(1-\sqrt{g_1})^2}\leq S_2< 1\nonumber\\
&~~~~~~~\frac{a(1-S_2)}{(1-\sqrt{g_2})^2}\leq S_1< 1\nonumber\\
&~~~~~~~0< [g_1,g_2].\nonumber
\end{IEEEeqnarray}

In a similar fashion, one can introduce two other optimization
problems, say $\tilde{W}_1(\mu)$ and $\tilde{W}_2(\mu)$, to obtain
upper bounds on $\sigma_{\mathscr{C}}(1,\mu)$ by using the upper
bounds on the support functions of channels in Class A2 and Class B.

\begin{theorem}[New Outer Bound]
For any rate pair $(R_1,R_2)$ achievable for the two-user weak Gaussian IC, the inequalities
\begin{IEEEeqnarray}{rl}
\mu_1 R_1+R_2\leq W(\mu_1)=\min\{W_1(\mu_1),W_2(\mu_1)\},\\
R_1+\mu_2 R_2\leq
\tilde{W}(\mu_2)=\min\{\tilde{W}_1(\mu_2),\tilde{W}_2(\mu_2)\},
\end{IEEEeqnarray}
hold for all $1\leq \mu_1,\mu_2$.
\end{theorem}

To obtain an upper bound on the sum rate, we can apply the following
inequality:
\begin{equation}
\mathscr{C}_{\text{sum}}\leq \min_{1\leq \mu_1,\mu_2}
\frac{(\mu_2-1)W(\mu_1)+(\mu_1-1)\tilde{W}(\mu_2)}{\mu_1\mu_2-1}.
\end{equation}

\subsection{Han-Kobayashi Achievable region}
In this sub-section, we aim at characterizing $\mathscr{G}$ for the
weak Gaussian IC. To this end, we first investigate some properties
of $\mathscr{G}_0(P_1,P_2,\alpha,\beta)$. First of all, we show that
none of the inequalities in describing $\mathscr{G}_0$ is redundant.
\begin{figure}
\centering \includegraphics[scale=0.75]{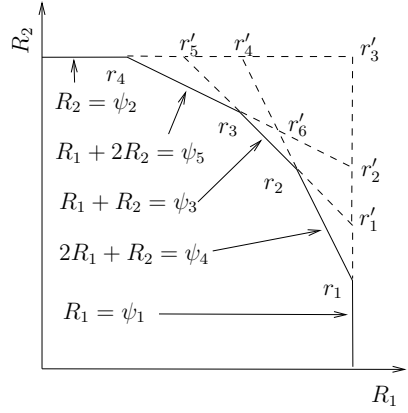}
\caption{$\mathscr{G}_0$ for the weak Gaussian IC. $r_1$, $r_2$,
$r_3$, and $r_4$ are extreme points of $\mathscr{G}_0$ in the
interior of the first quadrant.}\label{extreme points}
\end{figure}
In Figure \ref{extreme points}, all possible extreme points are
shown. It is easy to prove that $r'_i\notin \mathscr{G}_0$ for
$i\in\{1,2,\ldots,6\}$. For instance, we consider
$r'_6=\left(\frac{2\psi_4-\psi_5}{3},\frac{2\psi_5-\psi_4}{3}\right)$.
Since $\psi_{31}+\psi_{32}+\psi_{33}=\psi_4+\psi_5$ (see Section
II.C), we have
\begin{IEEEeqnarray}{rl}
\psi_3&=\min\{\psi_{31},\psi_{32},\psi_{33}\}\nonumber\\
&\leq \frac{1}{3}(\psi_{31}+\psi_{32}+\psi_{33})\nonumber\\
&=\frac{1}{3}(\psi_4+\psi_5).\nonumber
\end{IEEEeqnarray}
However, $\frac{1}{3}(\psi_4+\psi_5)$ is the sum of the components
of $r'_6$. Therefore, $r'_6$ violates (\ref{HK3}) in the definition
of the HK achievable region. Hence, $r'_6\notin \mathscr{G}_0$. As
another example, let us consider $r'_1=(\psi_1,\psi_3-\psi_1)$. We
claim that $r'_1$ violates (\ref{HK4}). To this end, we need to show
that $\psi_4\leq\psi_3+\psi_1$. However, it is easy to see that
$\psi_4\leq\psi_{31}+\psi_1$, $\psi_4\leq\psi_{32}+\psi_1$, and
$\psi_4\leq\psi_{33}+\psi_1$ reduce to $0\leq
(1-\alpha)(1-b+\beta(1-ab)P_2)$, $0\leq (1-\beta)(1-a+(1-ab)P_1)$,
and $0\leq (1-\alpha)(1-\beta)aP_2$, respectively. Therefore,
$r'_1\notin \mathscr{G}_0$.

We conclude that $\mathscr{G}$ has four extreme points in the
interior of the first quadrant, namely
\begin{IEEEeqnarray}{rl}
r_1&=(\psi_1,\psi_4-2\psi_1),\\
r_2&=(\psi_4-\psi_3,2\psi_3-\psi_4),\\
r_3&=(2\psi_3-\psi_5,\psi_5-\psi_3),\\
r_4&=(\psi_5-2\psi_2,\psi_2).
\end{IEEEeqnarray}

Most importantly, $\mathscr{G}_0$ possesses the unique minimizer
property. To prove this, we need to show that $\hat{\mathbf{y}}$,
the minimizer of the optimization problem
\begin{IEEEeqnarray}{rl}\label{alaki4}
\sigma_{D_0}
(c_1,c_2,P_1,P_2,\alpha,\beta)&=\max\{c_1R_1+c_2R_2|A\mathbf{R}\leq
\Psi(P_1,P_2,\alpha,\beta)\}\nonumber\\
&=\min\{\mathbf{y}^t\Psi(P_1,P_2,\alpha,\beta)|A^t
\mathbf{y}=(c_1,c_2)^t,\mathbf{y}\geq0 \},
\end{IEEEeqnarray}
is independent of the parameters $P_1$, $P_2$, $\alpha$, and $\beta$
and only depends on $c_1$ and $c_2$. We first consider the case
$(c_1,c_2)=(\mu,1)$ for all $1\leq \mu$. It can be shown that for
$2<\mu$, the maximum of (\ref{alaki4}) is attained at $r_1$
regardless of $P_1$, $P_2$, $\alpha$, and $\beta$. Therefore, the
dual program has the minimizer $\hat{\mathbf{y}}=(\mu-2,0,0,1,0)^t$
which is clearly independent of $P_1$, $P_2$, $\alpha$, and $\beta$.
In this case, we have
\begin{equation}\label{alaki5}
\sigma_{D_0} (\mu,1,P_1,P_2,\alpha,\beta)=(\mu-2)\psi_1+\psi_4, ~2<
\mu.
\end{equation}
For $1\leq\mu\leq2$, one can show that $r_2$ and
$\hat{\mathbf{y}}=(0,0,2-\mu,\mu-1,0)^t$ are the maximizer and the
minimizer of (\ref{alaki4}), respectively. In this case, we have
\begin{equation}\label{alaki6}
\sigma_{D_0}
(\mu,1,P_1,P_2,\alpha,\beta)=(2-\mu)\psi_3+(\mu-1)\psi_4,~1\leq\mu\leq2.
\end{equation}

Next, we consider the case $(c_1,c_2)=(1,\mu)$ for all $1\leq \mu$.
Again, it can be shown that for $2<\mu$ and $1\leq\mu\leq2$,
$\hat{\mathbf{y}}=(0,\mu-2,0,0,1)^t$ and
$\hat{\mathbf{y}}=(0,0,2-\mu,0,\mu-1)^t$ minimizes (\ref{alaki4}),
respectively. Hence, we have
\begin{IEEEeqnarray}{rll}
\sigma_{D_0} (1,\mu,P_1,P_2,\alpha,\beta)&=(\mu-2)\psi_2+\psi_5,&~\text{if}~ 2<\mu,\\
\sigma_{D_0}
(1,\mu,P_1,P_2,\alpha,\beta)&=(2-\mu)\psi_3+(\mu-1)\psi_5,&~\text{if}~
1\leq\mu\leq2.
\end{IEEEeqnarray}
We conclude that the solutions of the dual program are always
independent of $P_1$, $P_2$, $\alpha$, and $\beta$. Hence,
$\mathscr{G}_0$ possesses the unique minimizer property.

\begin{theorem}
For the two-user weak Gaussian IC, time-sharing and concavification
result in the same region. In other words, $\mathscr{G}$ can be
fully characterized by using TD/FD and allocating power over three
different dimensions.
\end{theorem}
\begin{proof}
Since $\mathscr{G}_0$ possesses the unique minimizer property, from
Theorem \ref{the unique minimizer property theorem}, we deduce that
$\mathscr{G}=\mathscr{G}_2$. Moreover, using Theorem
\ref{caratheodory}, the number of frequency bands is at most three.
\end{proof}

To obtain the support function of $\mathscr{G}_2$, we need to obtain
$g(c_1,c_2,P_1,P_2,\alpha,\beta)$ defined in (\ref{conv-3}). Since
$\mathscr{G}_0$ possesses the unique minimizer property,
(\ref{conv-3}) can be simplified. Let us consider the case where
$(c_1,c_2)=(\mu,1)$ for $\mu>2$. It can be shown that for this case
\begin{equation}
g=\max_{(\alpha,\beta)\in[0,1]^2} (\mu-2)\psi_1(P_{1},P_{2},\alpha,\beta)+ \psi_4(P_{1},P_{2},\alpha,\beta).
\end{equation}
Substituting into (\ref{conv-2}), we obtain
\begin{IEEEeqnarray}{rl}
\sigma_{\mathscr{G}_2} (\mu,1,P_1,P_2)=&\max \sum_{i=1}^{3} \lambda_i\left[ (\mu-2)\psi_1(P_{1i},P_{2i},\alpha_i,\beta_i)+\psi_4(P_{1i},P_{2i},\alpha_i,\beta_i)\right]\\
&\text{subject to:}\nonumber\\
&~~~~~~~\sum_{i=1}^{3} \lambda_i=1\nonumber\\
&~~~~~~~\sum_{i=1}^{3} \lambda_i P_{1i}\leq P_1\nonumber\\
&~~~~~~~\sum_{i=1}^{3} \lambda_i P_{2i}\leq P_2\nonumber\\
&~~~~~~~0\leq\lambda_i,0\leq P_{1i},0\leq P_{2i}, ~\forall i\in\{1,2,3\}\nonumber\\
&~~~~~~~0\leq\alpha_i\leq1,0\leq\beta_i\leq1, ~\forall
i\in\{1,2,3\}.\nonumber
\end{IEEEeqnarray}

\begin{figure}
\centering \includegraphics[scale=0.75]{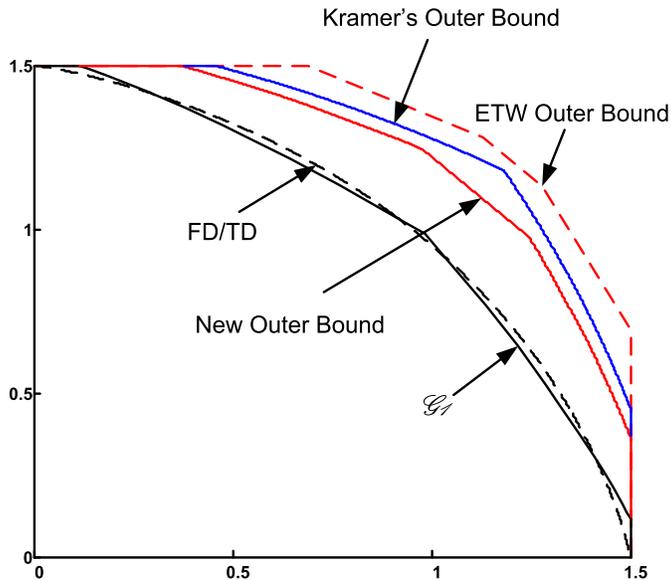}
\caption{Comparison between different bounds for the symmetric weak
Gaussian IC when $P=7$ and $a=0.2$.}\label{figure-weak-1}
\end{figure}

\begin{figure}
\centering \includegraphics[scale=0.75]{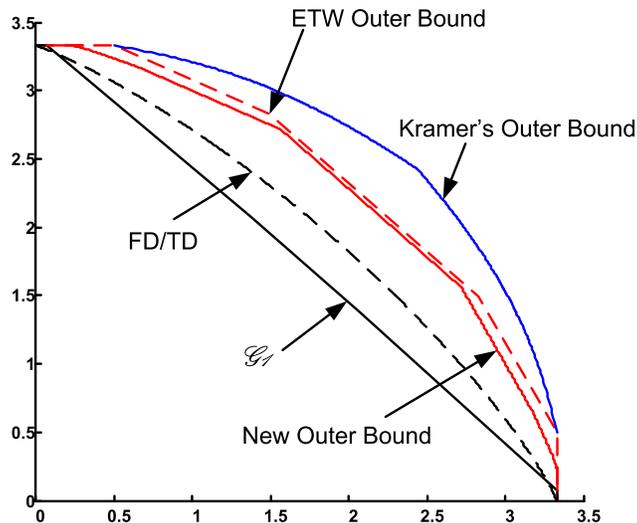}
\caption{Comparison between different bounds for the symmetric weak
Gaussian IC when $P=100$ and $a=0.1$.}\label{figure-weak-2}
\end{figure}

For other ranges of $(c_1,c_2)$, a similar optimization problem can
be formed. It is worth noting that even though the number of
parameters in characterizing $\mathscr{G}$ is reduced, it is still
prohibitively difficult to characterize boundary points of
$\mathscr{G}$. In Figures (\ref{figure-weak-1}) and
(\ref{figure-weak-2}), different bounds for the symmetric weak
Gaussian IC are plotted. As shown in these figures, the new outer
bound is tighter than the previously known bounds.

\section{One-sided Gaussian Interference Channels}
Throughout this section, we consider the one-sided Gaussian IC
obtained by setting $b=0$, i.e, the second receiver incurs no
interference from the first transmitter. One can further split the
class of one-sided ICs into two subclasses: the {\em strong
one-sided IC} and the {\em weak one-sided IC}. For the former,
$a\geq1$ and the capacity region is fully characterized
\cite{IC:SASON}. In this case, the capacity region is the union of
all rate pairs $(R_1,R_2)$ satisfying
\begin{IEEEeqnarray}{rl}
R_1&\leq \gamma{(P_1)},\nonumber\\
R_2&\leq \gamma(P_2),\nonumber\\
R_1+R_2&\leq \gamma(P_1+aP_2).\nonumber
\end{IEEEeqnarray}
For the latter, $a<1$ and the full characterization of the capacity
region is still an open problem. Therefore, we always assume $a<1$.
Three important results are proved for this channel. The first one,
proved by Costa in \cite{IC:COSTA}, states that the capacity region
of the weak one-sided IC is equivalent to that of the degraded IC
with an appropriate change of parameters. The second one, proved by
Sato in \cite{ZIC:SATO}, states that the capacity region of the
degraded Gaussian IC is outer bounded by the capacity region of a
certain degraded broadcast channel. The third one, proved by Sason
in \cite{IC:SASON}, characterizes the sum capacity by combining
Costa's and Sato's results.

In this section, we provide an alternative proof for the outer bound
obtained by Sato. We then characterize the full HK achievable region
where Gaussian codebooks are used, i.e., $\mathscr{G}$.

\subsection{Sum Capacity}
For the sake of completeness, we first state the sum capacity result
obtained by Sason.

\begin{theorem}[Sason]
The rate pair
$\left(\gamma\left(\frac{P_1}{1+aP_2}\right),\gamma(P_2)\right)$  is
an extreme point of the capacity region of the one-sided Gaussian
IC. Moreover, the sum capacity of the channel is attained at this
point.
\end{theorem}

\subsection{Outer Bound}
In \cite{ZIC:SATO}, Sato derived an outer bound on the capacity of
the degraded IC. This outer bound can be used for the weak one-sided
IC as well. This is due to Costa's result which states that the
capacity region of the degraded Gaussian IC is equivalent to that of
the weak one-sided IC with an appropriate change of parameters.

\begin{theorem}[Sato]
If the rate pair $(R_1,R_2)$ belongs to the capacity region of the weak one-sided IC, then it satisfies
\begin{equation}
\begin{array}{ll}
R_{1}&\leq \gamma{\left(\frac{(1-\beta)P}{1/a+\beta P}\right)},\\
R_{2}&\leq \gamma{(\beta P)},
\end{array}
\end{equation}
for all $\beta\in[0,1]$ where $P=P_1/a+P_2$.
\end{theorem}
\begin{proof}
Since the sum capacity is attained at the point where User 2
transmits at its maximum rate $R_2=\gamma(P_2)$, other boundary
points of the capacity region can be obtained by characterizing the
solutions of $\sigma_{\mathscr{C}}(\mu,1)=\max\left\{ \mu
R_1+R_2|(R_1,R_2)\in\mathscr{C}\right\}$ for all $1\leq \mu$. Using
Fano's inequality, we have
\begin{IEEEeqnarray}{rl}
n(\mu R_1+R_2)\stackrel{}{\leq}& \mu I(x_1^n;y_{1}^n)+I(x_2^n;y_{2}^n)+n\epsilon_n\nonumber\\
\stackrel{}{=}& \mu h(y_{1}^n)-\mu h(y_{1}^n|x_1^n) +h(y_2^n)-h(y_{2}^n|x_2^n)+n\epsilon_n\nonumber\\
\stackrel{}{=}& [\mu h(x_1^n+\sqrt{a}x_2^n+z_1^n)-h(z_2^n)]+[h(x_2^n+z_2^n)-\mu h(\sqrt{a}x_2^n+z_1^n)]+n\epsilon_n\nonumber\\
\stackrel{(a)}{\leq}& \frac{\mu n}{2}\log\left[2\pi e(P_1+aP_2+1)\right]-\frac{n}{2}\log(2\pi e)+[h(x_2^n+z_2^n)-\mu h(\sqrt{a}x_2^n+z_1^n)]+n\epsilon_n\nonumber\\
\stackrel{(b)}{\leq}& \frac{\mu n}{2}\log\left[2\pi
e(P_1+aP_2+1)\right]-\frac{n}{2}\log(2\pi
e)+nf_h(P_2,1,1,a,\mu)+n\epsilon_n,\nonumber
\end{IEEEeqnarray}
where (a) follows from the fact that Gaussian distribution maximizes
the differential entropy for a given constraint on the covariance
matrix and (b) follows from the definition of $f_h$ in
(\ref{fh-function}).

Depending on the value of $\mu$, we consider the following two
cases:

1- For $1\leq \mu \leq \frac{P_2+1/a}{P_2+1}$, we have
\begin{equation}\label{alaki61}
\mu R_1+R_2\leq\mu\gamma\left(\frac{P_1}{1+aP_2}\right)+\gamma(P_2).
\end{equation}
In fact, the point
$\left(\gamma\left(\frac{P_1}{1+aP_2}\right),\gamma(P_2)\right)$
which is achievable by treating interference as noise at Receiver 1,
satisfies (\ref{alaki61}) with equality. Therefore, it belongs to
the capacity region. Moreover, by setting $\mu=1$, we deduce that
this point corresponds to the sum capacity of the one-sided Gaussian
IC. This is in fact an alternative proof for Sason's result.

2- For $\frac{P_2+1/a}{P_2+1}< \mu \leq \frac{1}{a}$, we have
\begin{equation}
\mu R_1+R_2\leq
\frac{\mu}{2}\log\left(P_1+aP_2+1\right)+\frac{1}{2}\log\left(
\frac{1/a-1}{\mu-1}\right)-\frac{\mu}{2}\log\left(\frac{a\mu
(1/a-1)}{\mu-1}\right).
\end{equation}
Equivalently, we have
\begin{equation}\label{one-sided-outer}
\mu R_1+R_2\leq
\frac{\mu}{2}\log\left(\frac{(aP+1)(\mu-1)}{\mu(1-a)}\right)+\frac{1}{2}\log\left(
\frac{1/a-1}{\mu-1}\right),
\end{equation}
where $P=P_1/a+P_2$. Let us define $E_1$ as the set of all rate
pairs $(R_1,R_2)$ satisfying (\ref{one-sided-outer}), i.e.
\begin{equation}\label{alaki7}
E_1=\left\{(R_1,R_2)|\mu R_1+R_2\leq
\frac{\mu}{2}\log\left(\frac{(aP+1)(\mu-1)}{\mu(1-a)}\right)+\frac{1}{2}\log\left(
\frac{1/a-1}{\mu-1}\right),~\forall \frac{P_2+1/a}{P_2+1}< \mu \leq
\frac{1}{a}\right\}.
\end{equation}
We claim that $E_1$ is the dual representation of the region defined
in the statement of the theorem, see (\ref{dual representation}). To
this end, we define $E_2$ as
\begin{equation}
E_2=\left\{(R_1,R_2)|R_{1}\leq
\gamma{\left(\frac{(1-\beta)P}{1/a+\beta P}\right)}, R_{2}\leq
\gamma{(\beta P)},~\forall \beta\in[0,1]\right\}
\end{equation}
We evaluate the support function of $E_2$ as
\begin{equation}
\sigma_{E_2}(\mu,1)=\max\left\{\mu R_1+R_2|(R_{1}, R_{2})\in
E_2\right\}.
\end{equation}
It is easy to show that $\beta=\frac{1/a-1}{P(\mu-1)}$ maximizes the
above optimization problem. Therefore, we have
\begin{equation}
\sigma_{E_2}(\mu,1)=\frac{\mu}{2}\log\left(\frac{(aP+1)(\mu-1)}{\mu(1-a)}\right)+\frac{1}{2}\log\left(
\frac{1/a-1}{\mu-1}\right).
\end{equation}
Since $E_2$ is a closed convex set, we can use (\ref{dual
representation}) to obtain its dual representation which is indeed
equivalent to (\ref{alaki7}). This completes the proof.
\end{proof}

\subsection{Han-Kobayashi Achievable Region}
In this subsection, we characterize $\mathscr{G}_0$,
$\mathscr{G}_1$, $\mathscr{G}_2$, and $\mathscr{G}$ for the weak
one-sided Gaussian IC. $\mathscr{G}_0$ can be characterized as
follows. Since there is no link between Transmitter 1 and Receiver
2, User 1's message in the HK achievable region is only the private
message, i.e., $\alpha=1$. In this case, we have
\begin{IEEEeqnarray}{rl}
\psi_1&=\gamma\left(\frac{P_1}{1+a\beta P_2}\right),\\
\psi_2&=\gamma(P_2),\\
\psi_{31}&= \gamma\left(\frac{P_1+a(1-\beta)P_2}{1+a\beta P_2}\right)+\gamma(\beta P_2),\\
\psi_{32}&= \gamma\left(\frac{P_1}{1+a\beta P_2}\right)+\gamma(P_2),\\
\psi_{33}&= \gamma\left(\frac{P_1+a(1-\beta)P_2}{1+a\beta P_2}\right)+\gamma(\beta P_2),\\
\psi_4&= \gamma\left(\frac{P_1+a(1-\beta)P_2}{1+a\beta P_2}\right)+\gamma\left(\frac{P_1}{1+a\beta P_2}\right)+\gamma(\beta P_2),\\
\psi_5&=\gamma(\beta P_2)+\gamma(P_2)+\gamma\left(\frac{P_1+a(1-\beta)P_2}{1+a\beta P_2}\right),
\end{IEEEeqnarray}
It is easy to show that
$\psi_3=\min\{\psi_{31},\psi_{32},\psi_{33}\}=\psi_{31}$,
$\psi_{31}+\psi_1=\psi_4$, $\psi_{31}+\psi_2=\psi_5$. Hence,
$\mathscr{G}_0$ can be represented as all rate pairs $(R_1,R_2)$
satisfying
\begin{IEEEeqnarray}{rl}
\label{alaki8}R_1 &\leq \gamma\left(\frac{P_1}{1+a\beta P_2}\right),\\
\label{alaki9}R_2 &\leq  \gamma(P_2),\\
\label{alaki10}R_1+R_2 &\leq \gamma\left(\frac{P_1+a(1-\beta)P_2}{1+a\beta P_2}\right)+\gamma(\beta P_2).
\end{IEEEeqnarray}

We claim that $\mathscr{G}_2=\mathscr{G}$. To prove this, we need to
show that $\mathscr{G}_0$ possesses the unique minimizer property.
$\mathscr{G}_0$ is a pentagon with two extreme points in the
interior of the first quadrant, namely $r_1$ and $r_2$ where
\begin{IEEEeqnarray}{rl}
r_1&=\left(\gamma\left(\frac{P_1}{1+a\beta P_2}\right),
\gamma{\left(\frac{(1-\beta)aP_{2}}{1+P_{1}+\beta aP_{2}}\right)}
+\gamma(\beta P_2)\right),\label{on-sided-extreme1}\\
r_2&=\left(\gamma\left(\frac{P_1+a(1-\beta)P_2}{1+a\beta
P_2}\right)+\gamma(\beta
P_2)-\gamma(P_2),\gamma(P_2)\right).\label{on-sided-extreme2}
\end{IEEEeqnarray}
Using above, it can be verified that $\mathscr{G}_0$ possesses the
unique minimizer property.

Next, we can use the optimization problem in (\ref{conv-2}) to
obtain the support function of $\mathscr{G}$. However, we only need
to consider $(c_1,c_2)=(\mu,1)$ for $\mu>1$. Therefore, we have
\begin{equation}
g(\mu,1,P_1,P_2,\beta)=\max_{0\leq\beta\leq1} \mu
\gamma{\left(\frac{P_{1}}{1+\beta aP_{2}}\right)} +\gamma{(\beta
P_{2})}+\gamma{\left(\frac{(1-\beta)aP_{2}}{1+P_{1}+\beta
aP_{2}}\right)}.
\end{equation}
Substituting into (\ref{conv-2}), we conclude that boundary points
of $\mathscr{G}$  can be characterized by solving the following
optimization problem:
\begin{IEEEeqnarray}{rl}
W=&\max \sum_{i=1}^3 \lambda_i\left[ \mu
\gamma{\left(\frac{P_{1i}}{1+\beta_i a P_{2i}}\right)}
+\gamma{(\beta_i
P_{2i})}+\gamma{\left(\frac{(1-\beta_i)aP_{2i}}{1+P_{1i}+\beta_i
aP_{2i}}\right)}\right]\\
&\text{subject to:}\nonumber\\
&~~~~~~~\sum_{i=1}^3 \lambda_i=1\nonumber\\
&~~~~~~~\sum_{i=1}^3 \lambda_i P_{1i}\leq P_1\nonumber\\
&~~~~~~~\sum_{i=1}^3 \lambda_i P_{2i}\leq P_2\nonumber\\
&~~~~~~~0\leq\beta_i\leq 1, ~\forall i\in\{1,2,3\}\nonumber\\
&~~~~~~~0\leq [P_{1i},P_{2i},\lambda_i],~\forall
i\in\{1,2,3\}.\nonumber
\end{IEEEeqnarray}

For the sake of completeness, we provide a simple description for
$\mathscr{G}_1$ in the next lemma.
\begin{lemma}\label{lem one-sided}
The region $\mathscr{G}_1$ can be represented as the collection of
all rate pairs $(R_1,R_2)$ satisfying
\begin{IEEEeqnarray}{rl}\label{bibiki}
R_{1}&\leq \gamma{\left(\frac{P_{1}}{1+a\beta' P_{2}}\right)},\\
R_{2}&\leq \gamma{(\beta'
P_{2})}+\gamma{\left(\frac{a(1-\beta')P_{2}}{1+P_{1}+a\beta'
P_{2}}\right)},\label{bibiki1}
\end{IEEEeqnarray}
for all $\beta'\in[0,1]$. Moreover, $\mathscr{G}_1$ is convex and
any point that lies on its boundary can be achieved by using
superposition coding and successive decoding.
\end{lemma}

\begin{proof}
Let $E$ denote the set defined in the above lemma. It is easy to
show that $E$ is convex and $E\subseteq\mathscr{G}_1$. To prove the
inverse inclusion, it suffices to show that the extreme points of
$\mathscr{G}_0$, $r_1$ and $r_2$ (see (\ref{on-sided-extreme1}) and
(\ref{on-sided-extreme2})) are inside $E$ for all $\beta\in[0,1]$.
By setting $\beta'=\beta$, we see that $r_1\in E$. To prove $r_2\in
E$, we set $\beta'=1$. We conclude that $r_2\in E$ if the following
inequality holds
\begin{equation}\label{inequality-test}
\gamma\left(\frac{P_1+a(1-\beta)P_2}{1+a\beta
P_2}\right)+\gamma(\beta P_2)-\gamma(P_2)\leq
\gamma{\left(\frac{P_{1}}{1+a P_{2}}\right)},
\end{equation}
for all $\beta\in[0,1]$. However, (\ref{inequality-test}) reduces to
$0\leq (1-a)(1-\beta)P_2$ which holds for all $\beta\in[0,1]$.
Hence, $\mathscr{G}_1\subseteq E$. Using these facts, it is
straightforward to show that the boundary points $\mathscr{G}_1$ are
achievable by using superposition coding and successive decoding.
\end{proof}

\begin{figure}
\centering \includegraphics[scale=0.75]{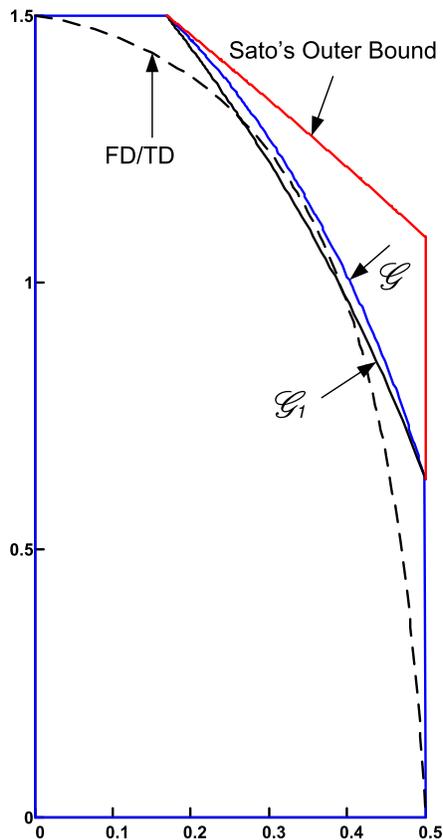}
\caption{Comparison between different bounds for the one-sided
Gaussian IC when $P_1=1$, $P_2=7$, and
$a=0.4$.}\label{figure-one-sided}
\end{figure}

Figure \ref{figure-one-sided} compares different bounds for the
one-sided Gaussian IC.

\section{Mixed Gaussian Interference Channels}
In this section, we focus on the mixed Gaussian Interference
channel. We first characterize the sum capacity of this channel.
Then, we provide an outer bound on the capacity region. Finally, we
investigate the HK achievable region. Without loss of generality, we
assume $a<1$ and $b\geq1$.

\subsection{Sum Capacity}
\begin{theorem}
The sum capacity of the mixed Gaussian IC with $a<1$ and $b\geq1$
can be stated as
\begin{equation}
\mathscr{C}_{sum}=\gamma\left(P_2\right)+\min\left\{\gamma\left(\frac{P_1}{1+aP_2}\right),
\gamma\left(\frac{bP_1}{1+P_2}\right)\right\}.
\end{equation}
\end{theorem}
\begin{proof}
We need to prove the achievability and converse for the theorem.

\textbf{Achievability part}: Transmitter 1 sends a common message to
both receivers, while the first user's signal is considered as noise
at both receivers. In this case, the rate
\begin{equation}\label{mixed-sum-1}
R_1=\min\left\{\gamma\left(\frac{P_1}{1+aP_2}\right),
\gamma\left(\frac{bP_1}{1+P_2}\right)\right\}
\end{equation}
is achievable. At Receiver 2, the signal from Transmitter 1 can be
decoded and removed. Therefore, User 2 is left with a channel
without interference and it can communicate at its maximum rate
which is
\begin{equation}\label{mixed-sum-2}
R_2=\gamma(P_2).
\end{equation}
By adding (\ref{mixed-sum-1}) and (\ref{mixed-sum-2}), we obtain the desired result.

\textbf{Converse part}: The sum capacity of the Gaussian IC is upper
bounded by that of the two underlying one-sided Gaussian ICs. Hence,
we can obtain two upper bounds on the sum rate. We first remove the
interfering link between Transmitter 1 and Receiver 2. In this case,
we have a one-sided Gaussian IC with weak interference. The sum
capacity of this channel is known \cite{IC:SASON}. Hence, we have
\begin{equation}\label{mixed-sum-3}
\mathscr{C}_{sum}\leq \gamma(P_2)+\gamma\left(\frac{P_1}{1+aP_2}\right).
\end{equation}
By removing the interfering link between Transmitter 2 and Receiver
1, we obtain a one-sided Gaussian IC with strong interference. The
sum capacity of this channel is known. Hence, we have
\begin{equation}
\mathscr{C}_{sum}\leq \gamma\left(bP_1+P_2\right),
\end{equation}
which equivalently can be written as
\begin{equation}\label{mixed-sum-4}
\mathscr{C}_{sum}\leq \gamma(P_2)+\gamma\left(\frac{bP_1}{1+P_2}\right).
\end{equation}
By taking the minimum of the right hand sides of (\ref{mixed-sum-3})
and (\ref{mixed-sum-4}), we obtain
\begin{equation}
\mathscr{C}_{sum}\leq\gamma\left(P_2\right)+\min\left\{\gamma\left(\frac{P_1}{1+aP_2}\right),
\gamma\left(\frac{bP_1}{1+P_2}\right)\right\}.
\end{equation}
This completes the proof.
\end{proof}

\begin{remark}
In an independent work \cite{shang}, the sum capacity of the mixed
Gaussian IC is obtained for a certain range of parameters, whereas
in the above theorem, we characterize the sum capacity of this
channel for the entire range of its parameters (see also
\cite{abolfazl2}).
\end{remark}

By comparing $\gamma\left(\frac{P_1}{1+aP_2}\right)$ with
$\gamma\left(\frac{bP_1}{1+P_2}\right)$, we observe that if
$1+P_2\leq b+abP_2$, then the sum capacity corresponds to the sum
capacity of the one-sided weak Gaussian IC, whereas if $1+P_2>
b+abP_2$, then the sum capacity corresponds to the sum capacity of
the one-sided strong IC. Similar to the one-sided Gaussian IC, since
the sum capacity is attained at the point where User 2 transmits at
its maximum rate $R_2=\gamma(P_2)$, other boundary points of the
capacity region can be obtained by characterizing the solutions of
$\sigma_{\mathscr{C}}(\mu,1)=\max\left\{ \mu
R_1+R_2|(R_1,R_2)\in\mathscr{C}\right\}$ for all $1\leq \mu$.

\subsection{New Outer Bound}

The best outer bound to date, due to Etkin {\em et al.}
\cite{Etkin:IC}, is obtained by using the Genie aided technique.
This bound is the union of all rate pairs $(R_1,R_2)$ satisfying
\begin{IEEEeqnarray}{rl}
R_1&\leq \gamma(P_1),\\
R_2&\leq \gamma(P_2),\\
R_1+R_2&\leq \gamma(P_2)+\gamma\left(\frac{P_1}{1+aP_2}\right),\\
R_1+R_2&\leq \gamma(P_2+bP_1),\\
2R_1+R_2&\leq
\gamma(P_1+aP_2)+\gamma\left(bP_1+\frac{P_2}{1+aP_2}\right)+
\gamma\left(\frac{P_1}{1+bP_1}\right).
\end{IEEEeqnarray}

The capacity region of the mixed Gaussian IC is inside the
intersection of the capacity regions of the two underlying one-sided
Gaussian ICs. Removing the link between Transmitter 1 and Receiver 2
results in a weak one-sided Gaussian IC whose outer bound $E_1$ is
the collection of all rate pairs $(R_1,R_2)$ satisfying
\begin{IEEEeqnarray}{rl}
R_1&\leq \gamma\left(\frac{(1-\beta)P'}{\beta P'+1/a}\right),\\
R_2&\leq \gamma(\beta P'),
\end{IEEEeqnarray}
for all $\beta\in[0,\beta_{\text{max}}]$, where $P'=P_1/a+P_2$ and
$\beta_{\text{max}}=\frac{P_2}{P'(1+P_1)}$. On the other hand,
removing the link between Transmitter 2 and Receiver 1 results in a
strong one-sided Gaussian IC whose capacity region $E_2$ is fully
characterized as the collection of all rate pairs $(R_1,R_2)$
satisfying
\begin{IEEEeqnarray}{rl}
R_1&\leq \gamma(b P_1),\\
R_2&\leq \gamma\left(P_2\right),\\
R_1+R_2&\leq \gamma(bP_1+P_2).
\end{IEEEeqnarray}

Using the channels in Class C, we upper bound
$\sigma_{\mathscr{C}}(\mu,1)$  based on the following optimization
problem:
\begin{IEEEeqnarray}{rl}
W(\mu)=&\min \frac{\mu-1}{2}\log\left(2\pi e(P_1+aP_2+1)\right)+\frac{1}{2}\log\left(2\pi e \left(\frac{P_2N_{22}}{P_2+N_{22}}+b'P_1+N_{21}\right)\right)\\
&~~~~~~~-\frac{1}{2}\log(2\pi e N_{21})-\frac{1}{2}\log(2\pi e N_{22})+ f_h(P_2,N_{22},1,a,\mu-1)\nonumber\\
&\text{subject to:}\nonumber\\
&~~~~~~~b'g_2=b\nonumber\\
&~~~~~~~b'\geq N_{21}\nonumber\\
&~~~~~~~aN_{22}\leq 1\nonumber\\
&~~~~~~~(1-\sqrt{g_2})^2N_{22}+g_2N_{21}=1\nonumber\\
&~~~~~~~0\leq [b',g_2,N_{22},N_{21}].\nonumber
\end{IEEEeqnarray}
By substituting $S=g_2N_{21}$, we obtain
\begin{IEEEeqnarray}{rl}
W(\mu)=&\min \frac{\mu-1}{2}\log\left(2\pi e(P_1+aP_2+1)\right)+\frac{1}{2}\log\left(2\pi e \left(\frac{P_2(1-S)}{(1-\sqrt{g_2})^2P_2+1-S}+\frac{bP_1+S}{g_2}\right)\right)\\
&~~~~~~~-\frac{1}{2}\log\left(\frac{2\pi e S}{g_2}\right)-\frac{1}{2}\log\left(\frac{2\pi e (1-S)}{(1-\sqrt{g_2})^2}\right)+ f_h\left(P_2,\frac{1-S}{(1-\sqrt{g_2})^2},1,a,\mu-1\right)\nonumber\\
&\text{subject to:}\nonumber\\
&~~~~~~~S<1\nonumber\\
&~~~~~~~a(1-S)\leq (1-\sqrt{g_2})^2\nonumber\\
&~~~~~~~0\leq [S,g_2].\nonumber
\end{IEEEeqnarray}

Hence, we have the following theorem that provides an outer bound on
the capacity region of the mixed Gaussian IC.

\begin{theorem}
For any rate pair $(R_1,R_2)$ achievable for the two-user mixed
Gaussian IC, $(R_1,R_2)\in E_1\bigcap E_2$. Moreover, the inequality
\begin{equation}
\mu R_1+R_2\leq W(\mu)
\end{equation}
holds for all $1\leq \mu$.
\end{theorem}

\subsection{Han-Kobayashi Achievable Region}
In this subsection, we study the HK achievable region for the mixed
Gaussian IC. Since Receiver 2 can always decode the message of the
first user, User 1 associates all its power to the common message.
User 2, on the other hand, allocates $\beta P_2$ and $(1-\beta)P_2$
of its total power to its private and common messages, respectively,
where $\beta\in[0,1]$. Therefore, we have
\begin{IEEEeqnarray}{rl}
\label{HK-mixed1}\psi_1 &=  \gamma\left(\frac{P_1}{1+a\beta P_2}\right),\\
\label{HK-mixed2}\psi_2 &=  \gamma(P_2),\\
\label{HK-mixed3}\psi_{31} &= \gamma\left(\frac{P_1+a(1-\beta)P_2}{1+a\beta P_2}\right)+\gamma(\beta P_2),\\
\label{HK-mixed4}\psi_{32} &= \gamma(P_2+bP_1),\\
\label{HK-mixed5}\psi_{33} &= \gamma\left(\frac{a(1-\beta)P_2}{1+a\beta P_2}\right)+\gamma(\beta P_2+bP_1),\\
\label{HK-mixed6}\psi_4 &= \gamma\left(\frac{P_1+a(1-\beta)P_2}{1+a\beta P_2}\right)+\gamma(\beta P_2+bP_1),\\
\label{HK-mixed7}\psi_5 &= \gamma(\beta P_2)+  \gamma(P_2+bP_1)+ \gamma\left(\frac{a(1-\beta)P_2}{1+a\beta P_2}\right).
\end{IEEEeqnarray}

\begin{figure}
\centering \includegraphics[scale=0.75]{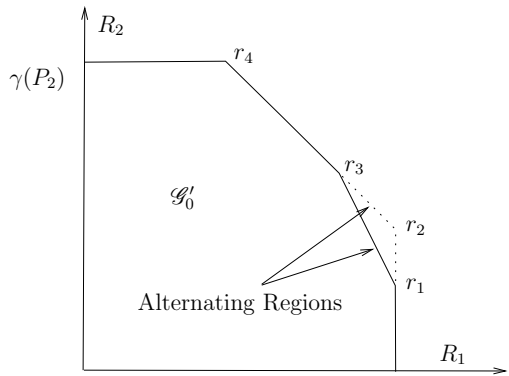}
\caption{The new region $\mathscr{G}'_0$ which is obtained by
enlarging $\mathscr{G}_0$.}\label{extreme points-mixed}
\end{figure}

Due to the fact that the sum capacity is attained at the point where
the second user transmits at its maximum rate, the last inequality
in the description of the HK achievable region can be removed.
Although the point $r'_5=(\psi_3-\gamma(P_2),\gamma(P_1))$ in Figure
\ref{extreme points} may not be in $\mathscr{G}_0$, this point is
always achievable due to the sum capacity result. Hence, we can
enlarge $\mathscr{G}_0$ by removing $r_3$ and $r_4$. Let us denote
the resulting region as $\mathscr{G}'_0$. Moreover, one can show
that $r'_2$, $r'_3$, $r'_4$, and $r'_6$ are still outside
$\mathscr{G}'_0$. However, for the mixed Gaussian IC, it is possible
that $r'_1$ belongs to $\mathscr{G}'_0$. In Figure \ref{extreme
points-mixed}, two alternative cases for the region $\mathscr{G}'_0$
along with the new labeling of its extreme points are plotted. The
new extreme points can be written as
\begin{IEEEeqnarray}{rl}
r_1 &=(\psi_1,\psi_4-2\psi_1),\nonumber\\
r_2 &=(\psi_1,\psi_3-\psi_1),\nonumber\\
r_3 &=(\psi_4-\psi_3,2\psi_3-\psi_4),\nonumber\\
r_4 &=(\psi_3-\psi_2,\psi_2).\nonumber
\end{IEEEeqnarray}
In fact, we have either $\mathscr{G}'_0=\text{conv}\{r_1,r_3,r_4\}$
or $\mathscr{G}'_0=\text{conv}\{r_2,r_4\}$.

To simplify the characterization of $\mathscr{G}_1$, we consider
three cases:
\begin{description}
\item[Case I:] ~~~$1+P_2\leq b+abP_2$.
\item[Case II:] ~~~$1+P_2> b+abP_2$ and $1-a\leq abP_1$.
\item[Case III:] ~~~$1+P_2> b+abP_2$ and $1-a> abP_1$.
\end{description}

\textbf{\underline{Case I ($1+P_2\leq b+abP_2$)}}: In this case,
$\psi_3=\psi_{31}$. Moreover, it is easy to verify that
$\psi_{31}+\psi_1\leq \psi_4$ which means (\ref{HK4}) is redundant
for the entire range of parameters. Hence,
$\mathscr{G}'_0=\text{conv}\{r_2,r_4\}$ consists of all rate pairs
$(R_1,R_2)$ satisfying
\begin{IEEEeqnarray}{rl}
\label{HK-mixed1c}R_1 &\leq  \gamma\left(\frac{P_1}{1+a\beta P_2}\right),\\
\label{HK-mixed2c}R_2 &\leq  \gamma\left(P_2\right),\\
\label{HK-mixed3c}R_1+R_2 &\leq  \gamma\left(\frac{P_1+a(1-\beta)P_2}{1+a\beta P_2}\right)+\gamma(\beta P_2),
\end{IEEEeqnarray}
where $\beta\in[0,1]$. Using a reasoning similar to the one used to
express boundary points of $\mathscr{G}_1$ for the one-sided
Gaussian IC, we can express boundary points of $\mathscr{G}_1$ as
\begin{IEEEeqnarray}{rl}
R_1 &\leq  \gamma\left(\frac{P_1}{1+a\beta P_2}\right),\\
R_{2} &\leq \gamma(\beta P_2)+
\gamma\left(\frac{a(1-\beta)P_2}{1+P_1+a\beta P_2}\right),
\end{IEEEeqnarray}
for all $\beta\in[0,1]$.
\begin{theorem}
For the mixed Gaussian IC satisfying $1\leq ab$, region
$\mathscr{G}$ is equivalent to that of the one sided Gaussian IC
obtained from removing the interfering link between Transmitter 1
and Receiver 2.
\end{theorem}
\begin{proof}
If $1\leq ab$, then $1+P_2\leq b+abP_2$ holds for all $P_1$ and
$P_2$. Hence, $\mathscr{G}'_0(P_1,P_2,\beta)$ is a pentagon defined
by (\ref{HK-mixed1c}), (\ref{HK-mixed2c}), and (\ref{HK-mixed2c}).
Comparing with the corresponding region for the one-sided Gaussian
IC, we see that $\mathscr{G}'_0$ is equivalent to $\mathscr{G}_0$
obtained for the one-sided Gaussian IC. This directly implies that
$\mathscr{G}$ is the same for both channels.
\end{proof}

\underline{\textbf{Case II} ($1+P_2> b+abP_2$ and $1-a\leq abP_1$)}:
In this case, $\psi_3=\min\{\psi_{31},\psi_{32}\}$. It can be shown
that $\mathscr{G}_1$ is the union of three regions $E_1$, $E_2$, and
$E_3$, i.e, $\mathscr{G}_0=E_1\bigcup E_2\bigcup E_3$. Region $E_1$
is the union of all rate pairs $(R_1,R_2)$ satisfying
\begin{IEEEeqnarray}{rl}
R_1 &\leq  \gamma\left(\frac{P_1}{1+a\beta P_2}\right),\\
R_{2} &\leq \gamma(\beta P_2)+ \gamma\left(\frac{a(1-\beta)P_2}{1+P_1+a\beta P_2}\right).
\end{IEEEeqnarray}
for all $\beta\in[0,\frac{b-1}{(1-ab)P_2}]$. Region $E_2$ is the
union of all rate pairs $(R_1,R_2)$ satisfying
\begin{IEEEeqnarray}{rl}
R_1 &\leq  \gamma\left(\frac{bP_1}{1+\beta P_2}\right),\\
R_{2} &\leq \gamma\left(\frac{P_1+a(1-\beta)P_2}{1+a\beta P_2}\right)+\gamma(\beta P_2)-\gamma\left(\frac{bP_1}{1+\beta P_2}\right).
\end{IEEEeqnarray}
for all $\beta\in[\frac{b-1}{(1-ab)P_2},
\frac{(b-1)P_1+(1-a)P_2}{(1-ab)P_1P_2+(1-a)P_2}]$. Region $E_3$ is
the union of all rate pairs $(R_1,R_2)$ satisfying
\begin{IEEEeqnarray}{rl}
R_1 &\leq  \gamma\left(\frac{bP_1(1+\frac{(1-a b)P_1}{1-a})}{1+bP_1+P_2}\right),\\
R_{2} &\leq \gamma\left(P_2\right),\\
R_1+R_2&\leq \gamma(bP_1+P_2).
\end{IEEEeqnarray}

\underline{\textbf{Case III} ($1+P_2> b+abP_2$ and $1-a> abP_1$)}:
In this case, $\psi_3=\min\{\psi_{31},\psi_{32}\}$. Similar to Case
II, we have $\mathscr{G}_1=E_1\bigcup E_2\bigcup E_3$, where regions
$E_1$, $E_2$, and $E_3$ are defined as follows. Region $E_1$ is the
union of all rate pairs $(R_1,R_2)$ satisfying
\begin{IEEEeqnarray}{rl}
R_1 &\leq  \gamma\left(\frac{P_1}{1+a\beta P_2}\right),\\
R_{2} &\leq \gamma(\beta P_2)+ \gamma\left(\frac{a(1-\beta)P_2}{1+P_1+a\beta P_2}\right).
\end{IEEEeqnarray}
for all $\beta\in[0,\frac{b-1}{(1-ab)P_2}]$. Region $E_2$ is the
union of all rate pairs $(R_1,R_2)$ satisfying
\begin{IEEEeqnarray}{rl}
R_1 &\leq  \gamma\left(\frac{P_1}{1+a\beta P_2}\right),\\
R_{2} &\leq \gamma\left(\frac{a(1-\beta)P_2}{1+P_1+a\beta P_2}\right)+\gamma(\beta P_2+bP_1)-\gamma\left(\frac{P_1}{1+a\beta P_2}\right).
\end{IEEEeqnarray}
for all $\beta\in[\frac{b-1}{(1-ab)P_2},1]$. Region $E_3$ is the
union of all rate pairs $(R_1,R_2)$ satisfying
\begin{IEEEeqnarray}{rl}
R_1 &\leq  \gamma\left(\frac{P_1}{1+aP_2}\right),\\
R_{2} &\leq \gamma\left(P_2\right),\\
R_1+R_2&\leq \gamma(bP_1+P_2).
\end{IEEEeqnarray}

\begin{figure}
\centering \includegraphics[scale=0.75]{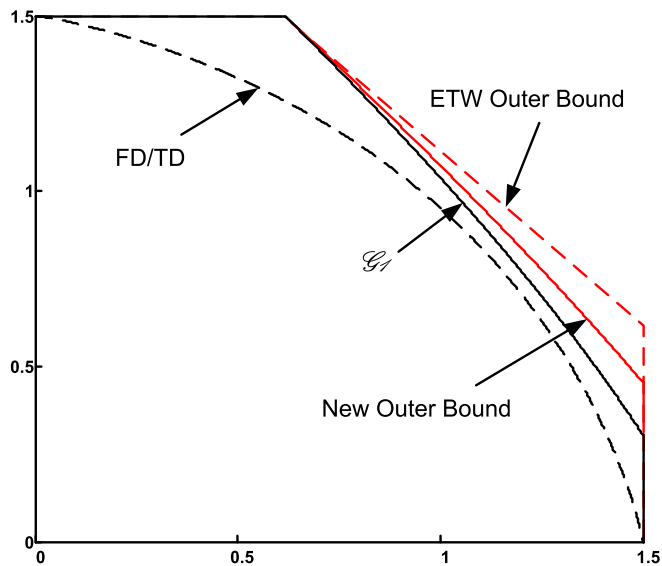}
\caption{Comparison between different bounds for the mixed Gaussian
IC when $1+P_2\leq b+abP_2$ (Case I) for $P_1=7$, $P_2=7$, $a=0.6$,
and $b=2$.}\label{figure-mixed-1}
\end{figure}

\begin{figure}
\centering \includegraphics[scale=0.75]{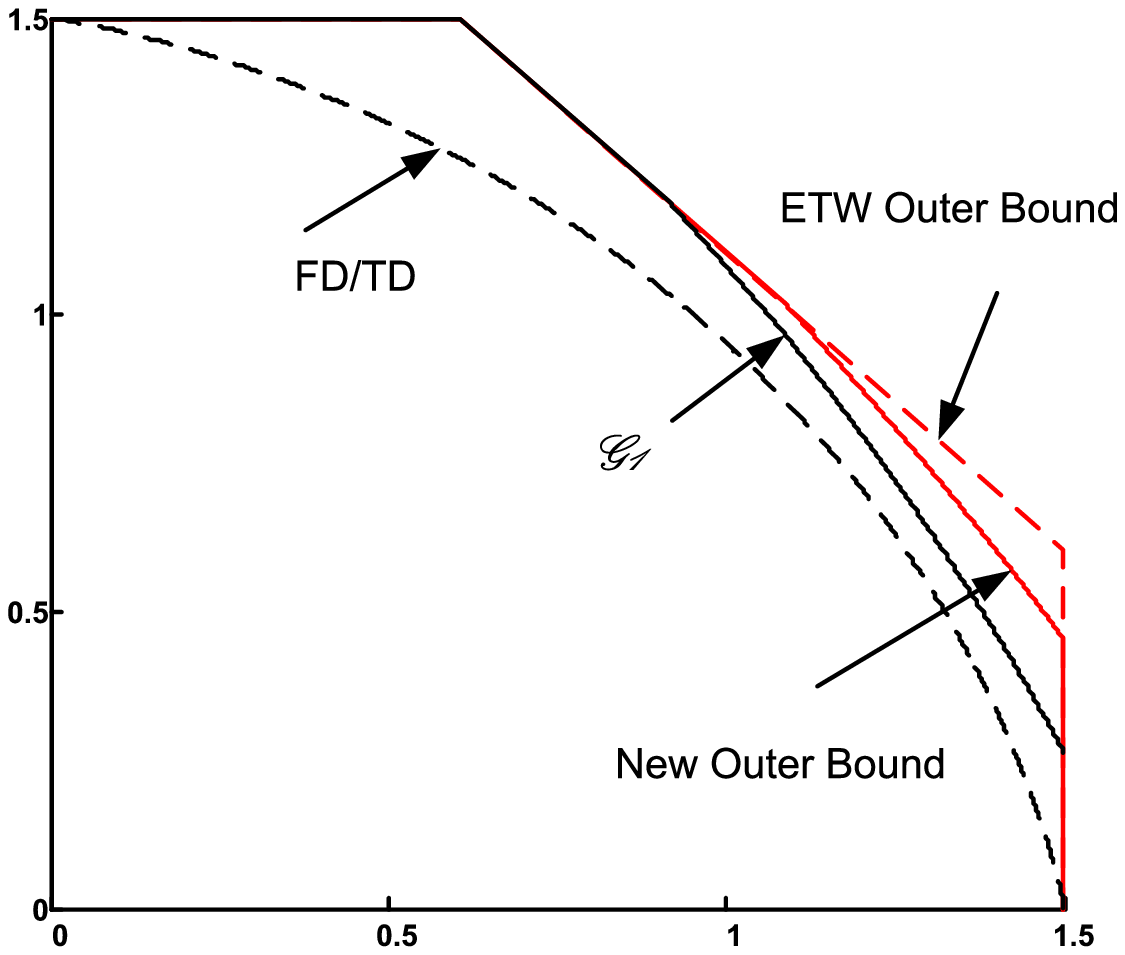}
\caption{Comparison between different bounds for the mixed Gaussian
IC when $1+P_2> b+abP_2$ and $1-a\leq abP_1$ (Case II) for $P_1=7$,
$P_2=7$, $a=0.4$, and $b=1.5$.}\label{figure-mixed-2}
\end{figure}

\begin{figure}
\centering \includegraphics[scale=0.75]{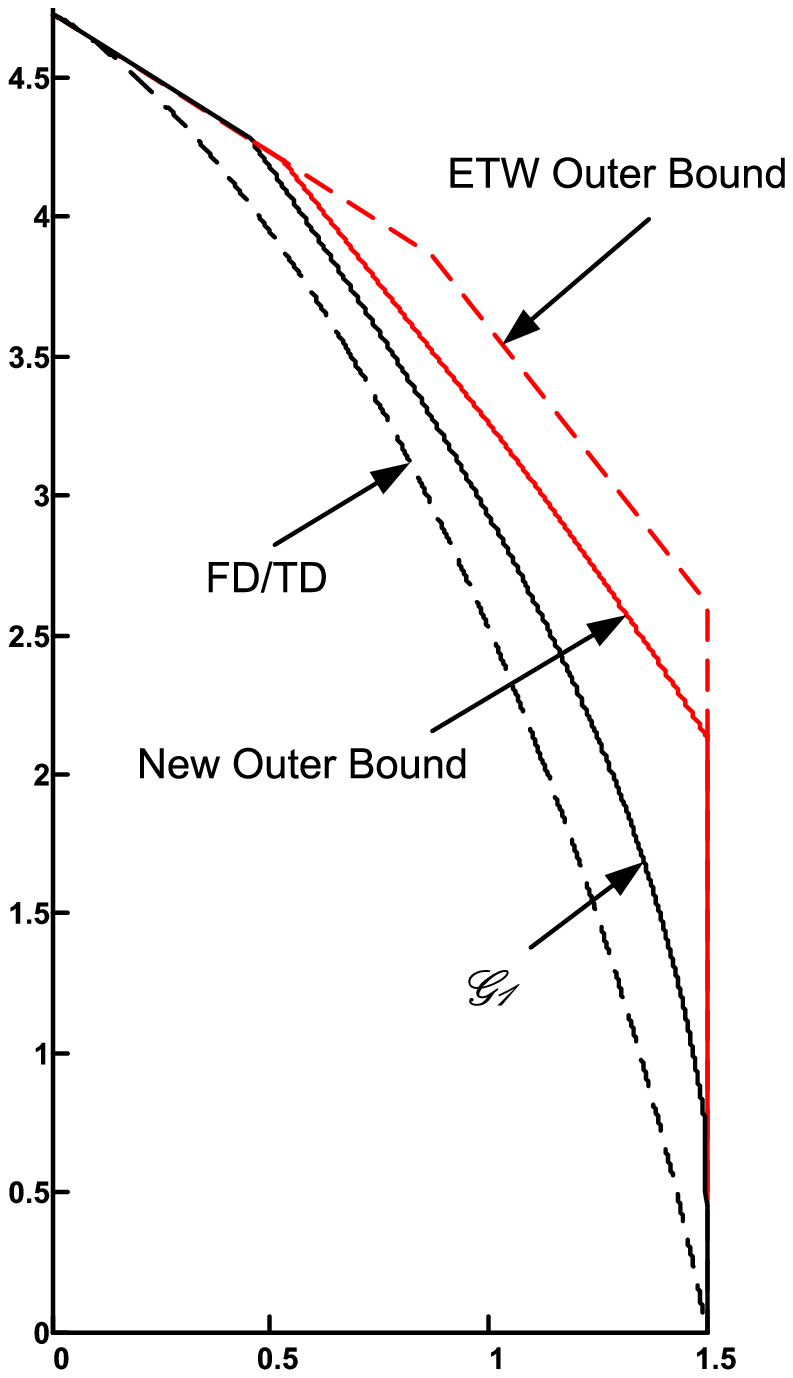}
\caption{Comparison between different bounds for the mixed Gaussian
IC when $1+P_2> b+abP_2$ and $1-a> abP_1$ (Case III) for $P_1=7$,
$P_2=700$, $a=0.01$, and $b=1.5$.}\label{figure-mixed-3}
\end{figure}

\begin{remark}
Region $E_3$ in Case II and Case III represents a facet that belongs
to the capacity region of the mixed Gaussian IC. It is important to
note that, surprisingly, this facet is obtainable when the second
transmitter uses both the common message and the private message.
\end{remark}

Different bounds are compared for the mixed Gaussian IC for Cases I,
II, and III in Figures \ref{figure-mixed-1}, \ref{figure-mixed-2},
and \ref{figure-mixed-3}, respectively.

\section{Conclusion}

We have studied the capacity region of the two-user Gaussian IC. The
sum capacities, inner bounds, and outer bounds have been considered
for three classes of channels: weak, one-sided, and mixed Gaussian
IC. We have used admissible channels as the main tool for deriving
outer bounds on the capacity regions.

For the weak Gaussian IC, we have derived the sum capacity for a
certain range of channel parameters. In this range, the sum capacity
is attained when Gaussian codebooks are used and interference is
treated as noise. Moreover, we have derived a new outer bound on the
capacity region. This outer bound is tighter than the Kramer's bound
and the ETW's bound. Regarding inner bounds, we have reduced the
computational complexity of the HK achievable region. In fact, we
have shown that when Gaussian codebooks are used, the full HK
achievable region can be obtained by using the naive HK achievable
scheme over three frequency bands.

For the one-sided Gaussian IC, we have presented an alternative
proof for the Sato's outer bound. We have also derived the full HK
achievable region when Gaussian codebooks are used.

For the mixed Gaussian IC, we have derived the sum capacity for the
entire range of its parameters. Moreover, we have presented a new
outer bound on the capacity region that outperforms ETW's bound. We
have proved that the full HK achievable region using Gaussian
codebooks is equivalent to that of the one-sided Gaussian IC for a
particular range of channel gains. We have also derived a facet that
belongs to the capacity region for a certain range of parameters.
Surprisingly, this facet is obtainable when one of the transmitters
uses both the common message and the private message.

\appendices

\bibliographystyle{IEEEtran}
\bibliography{sum}

\end{document}